\documentclass[a4paper,fleqn]{cas-dc}

\usepackage[utf8]{inputenc}

\usepackage{amsmath, amssymb}

\usepackage{geometry}
\usepackage{graphicx}

\usepackage{booktabs}
\usepackage{makecell}
\usepackage{enumitem}
\usepackage{tabularx}

\usepackage{hyperref}

\usepackage{siunitx}
\usepackage[gen]{eurosym}

\usepackage[intoc]{nomencl}
\makenomenclature

\usepackage[numbers]{natbib}

\begin{document}
\let\WriteBookmarks\relax
\def\floatpagepagefraction{1}
\def\textpagefraction{.001}

\shorttitle{A Robust Optimization Framework for Flexible Industrial Energy Scheduling}

\shortauthors{Sebastián Rojas-Innocenti et~al.}

\title [mode = title]{A Robust Optimization Framework for Flexible Industrial Energy Scheduling: Application to a Cement Plant with Market Participation}                      

\tnotetext[1]{A preprint of this work has been published on arXiv~\cite{rojasinnocenti2025robustoptimizationframeworkflexible}.}


%
\author[1]{Sebastián Rojas-Innocenti}[
                        prefix=Mr.,
                        orcid=0009-0009-4355-412X]

\cormark[1]



\credit{Conceptualization, Data Curation, Formal analysis, Investigation, Methodology, Resources, Software, Visualization, Validation, Writing - Original Draft, Writing - Review \& Editing}

\author[2]{Enrique Baeyens}[prefix=Dr.,
   ]
\credit{Conceptualization, Formal analysis, Investigation, Methodology, Resources, Supervision, Validation, Visualization, Writing - Review \& Editing}

\author[3]{Fernando Frechoso}[
                        prefix=Dr.]
\credit{Resources, Supervision, Validation, Visualization, Writing - Review \& Editing}

\author[1]{Alejandro Martín-Crespo}[
                        prefix=Mr.]
\credit{Funding acquisition, Resources, Writing - Review \& Editing}

\author[1]{Sergio Saludes-Rodil}[
                        prefix=Dr.]
\credit{Funding acquisition, Project administration, Resources, Supervision, Writing - Review \& Editing}

\affiliation[1]{organization={CARTIF Technology Centre},
    city={Valladolid},
    country={Spain}}

\affiliation[2]{organization={Instituto de las Tecnologías Avanzadas de la Producción, Valladolid University},
    city={Valladolid},
    country={Spain}}

\affiliation[3]{organization={Departamento de Ingeniería Eléctrica, Valladolid University},
    city={Valladolid},
    country={Spain}}

\cortext[cor1]{Corresponding author}



\begin{abstract}
This paper presents a scenario based robust optimization framework for short term energy scheduling in electricity intensive industrial plants, explicitly addressing uncertainty in planning decisions. The model is formulated as a two-stage Mixed Integer Linear Program (MILP) and integrates a hybrid scenario generation method capable of representing uncertain inputs such as electricity prices, renewable generation, and internal demand.
A convex objective function combining expected and worst case operational costs allows for tunable risk aversion, enabling planners to balance economic performance and robustness. The resulting schedule ensures feasibility across all scenarios and supports coordinated use of industrial flexibility assets, including battery energy storage and shiftable production.
To isolate the effects of market volatility, the framework is applied to a real world cement manufacturing case study considering only day-ahead electricity price uncertainty, with all other inputs treated deterministically. Results show improved resilience to forecast deviations, reduced cost variability, and more consistent operations. The proposed method offers a scalable and risk-aware approach for industrial flexibility planning under uncertainty.
\end{abstract}




\maketitle

\section{Introduction}

The increasing penetration of variable renewable energy sources has introduced significant volatility into electricity markets, posing new challenges for electricity-intensive industrial consumers. These facilities are now expected to deliver demand-side flexibility—not only to reduce energy costs but also to support grid stability. However, effective participation in electricity markets requires decisions to be made under uncertainty, particularly in short-term operations, where both prices and internal demand may deviate from forecasts.

Prior studies have explored industrial demand-side flexibility primarily through deterministic optimization models~\cite{rojasinnocenti2024electrical, rojas_comparative}, which assume perfect foresight of electricity prices and production needs. While valuable for baseline assessments, these models may underperform in practice due to their inability to cope with the high uncertainty present in volatile markets or variable-load processes.

To address this limitation, we propose a robust optimization framework formulated as a two-stage Mixed Integer Linear Program (MILP). The model explicitly accounts for multiple sources of uncertainty—including day-ahead electricity prices, renewable generation, and internal demand—via a hybrid, data-driven scenario generation approach. The objective function is defined as a convex combination of expected and worst-case operational costs, enabling decision-makers to tune the risk aversion level through a continuous parameter. The resulting solution is guaranteed to be feasible under all considered scenarios and coordinates the operation of industrial flexibility assets such as battery energy storage systems (BESS) and shiftable production processes.

The framework is applied to a real-world cement manufacturing plant in Spain. Although the model can handle multiple uncertainty sources, the case study focuses specifically on price uncertainty to isolate its effect on scheduling and flexibility activation.

\textbf{Structure of the paper:} Section~\ref{sec:literature} reviews relevant literature on industrial flexibility and robust optimization. Section~\ref{sec:Methodology} presents the proposed methodology, including the scenario generation and MILP formulation. Section~\ref{sec:case_study} describes the industrial use case. Section~\ref{sec:results} provides simulation results, and Section~\ref{sec:conclusions} concludes with key findings and future research directions.

\section{Literature Review}
\label{sec:literature}

This section reviews two complementary strands of literature that form the foundation for the proposed robust scheduling framework: Deterministic optimization approaches for industrial demand response, and robust optimization methods for energy systems under uncertainty.

\subsection*{Deterministic Demand Response in Industry}

Industrial demand-side response (DSR) is increasingly seen as a key enabler of power system flexibility~\cite{rollert_demand_2022, pierri_integrated_2020}. Deterministic optimization techniques—such as MILP, heuristics, and Model Predictive Control (MPC)—are commonly used to align electricity consumption with price signals while maintaining process feasibility~\cite{boldrini_demand_2023, Zhao_2014}.

In the cement industry, flexibility is primarily concentrated in stages like raw milling and grinding, which are decoupled from the kiln through intermediate storage~\cite{lee_evaluation_2020, olsen_opportunities_2011, rombouts_flexible_2021}. For instance, Rombouts et al.~\cite{rombouts_flexible_2023} estimate that around 24\% of electricity usage in a Belgian cement plant is shiftable. Other examples include the use of genetic algorithms to reduce energy costs under time-of-use tariffs~\cite{parejo_guzman_methodological_2022}, and MILP-based scheduling to participate in ancillary service markets~\cite{zhang_mpc_2023, tong_pv_bess_2024}.

Similar flexibility is observed in the steel industry, where electric arc furnaces (EAFs) operate in batches and tolerate scheduling shifts~\cite{paulus_potential_2011}. Boldrini et al.~\cite{boldrini_demand_2023} propose a MILP model for joint participation in day-ahead and reserve markets, achieving both cost savings and revenue improvements.

Other sectors—such as pulp and paper~\cite{arias2022demand}, water treatment~\cite{torregrossa2016energy}, and food processing~\cite{rahmani_residential_2023}—also exploit flexibility through storage or interruptible loads. Recent meta-analyses show that industrial sites with BESS outperform residential systems in up-regulation services~\cite{adiguzel_global_2024}.

Overall, these works confirm the technical and economic potential of industrial DSR. However, their reliance on perfect foresight restricts their effectiveness in real-world settings marked by uncertainty.

\subsection*{Robust Optimization under Uncertainty}

Robust optimization provides a viable alternative to deterministic models for planning under uncertainty. By ensuring feasibility across a range of possible realizations, these models enhance resilience against forecast deviations~\cite{ben2009robust, shapiro2009lectures, bertsimas2011theory, Conejo2010}.

For example, Fang et al.~\cite{fang2022robust} develop robust schedules for industrial loads under price uncertainty, outperforming both deterministic and stochastic models. Other applications include building control~\cite{parisio2016stochastic}, microgrids~\cite{oldewurtel2010use}.

More recent studies have extended robust optimization to complex industrial systems. Yang et al.~\cite{yang2024two} introduce a two-stage framework for wind uncertainty and demand response in large consumers. Xu et al.~\cite{xu2024improved} propose a distributionally robust model for spatiotemporal renewable uncertainty with embedded demand response.

These contributions reflect a growing shift toward risk-aware, uncertainty-resilient industrial energy management.

\subsection*{Research Gaps and Contributions of This Study}

Despite the progress outlined above, several important research gaps remain—particularly in the cement sector. First, few studies jointly optimize production scheduling and flexibility activation under realistic, scenario based uncertainty. Most existing models rely on deterministic formulations or limited uncertainty representations, which fail to capture the variability inherent in real industrial environments. Second, there is a lack of models that incorporate a tunable risk aversion mechanism, limiting the ability to balance expected cost and robustness in a systematic way. Finally, only a handful of works consider the integration of market based flexibility mechanisms such as the continuous intraday market (SIDC), or the simultaneous use of BESS alongside process level flexibility.

This study addresses these gaps by proposing a scenario based robust MILP framework that enables explicit risk tuning through a convex combination of worst case and expected costs. The model embeds key industrial constraints—including binary scheduling decisions, inventory management, and detailed BESS operation—within a unified and computationally tractable formulation. To demonstrate practical relevance, the framework is validated through a real world case study in cement manufacturing, using empirical market data and representative operational parameters.

In doing so, the paper contributes a practical, modular, and industry ready tool for robust energy scheduling in uncertainty prone industrial environments.

\section{Methodology}
\label{sec:Methodology}

This section presents the integrated methodological framework for flexible industrial production scheduling under uncertainty. The approach combines a hybrid scenario generation technique with a two-stage robust Mixed Integer Linear Programming model, extending previously validated two-stage formulations with explicit uncertainty modeling.

\subsection{Methodological Overview}

The methodology integrates two main components:

\begin{itemize}[left=0pt]
    \item \textbf{Hybrid Scenario Generation}: A data driven procedure combining time series forecasting, residual based stochastic simulation, and clustering, to produce representative scenarios of uncertain parameters.
    \item \textbf{Two-Stage Robust Optimization}: A MILP framework for production scheduling and flexibility activation under uncertainty, formulated over the generated scenario set.
\end{itemize}

The two-stage structure is preserved across deterministic and robust models, with uncertainty treatment being the primary differentiator.

\subsection{Hybrid Scenario Generation for Robust Optimization}
\label{sec:scenario_generation}

To explicitly incorporate uncertainty into the optimization model, a hybrid scenario generation technique is adopted. The process consists of three sequential phases designed to generate a compact yet expressive set of scenarios that represent the variability of key inputs:

\begin{enumerate}[left=0pt]
    \item \textbf{Forecasting Phase:}  
    Historical time series of uncertain variables—including day-ahead electricity prices, photovoltaic (PV) generation, and internal production demand—are modeled using seasonal ARIMA techniques~\cite{hyndman2018forecasting}. These models capture key features such as autocorrelation, trend, and seasonality. The resulting deterministic forecasts serve as a reference trajectory anchoring the scenario generation process.
    
    \item \textbf{Residual Simulation:}  
    To reflect forecast uncertainty, residuals from the fitted ARIMA models are resampled and superimposed on the deterministic forecast. This procedure generates a large ensemble of stochastic trajectories that retain temporal coherence while incorporating variability consistent with historical forecast errors.
    
    \item \textbf{Clustering and Scenario Reduction:}  
    The ensemble is normalized to remove magnitude biases and clustered using the K-Means algorithm~\cite{lloyd1982least}. The centroids of each cluster define the representative scenarios. Probabilities are assigned based on the relative size of each cluster, enabling the construction of a weighted scenario set for expected value calculations.
\end{enumerate}

This three-phase structure encapsulates the key strengths of the hybrid scenario generation method. Table~\ref{tab:features_scenario_generation} summarizes its main design features and advantages, highlighting how each step contributes to building a compact, representative, and risk-aware scenario set suitable for robust optimization.

\begin{table}[h]
\centering
\caption{Key features of the hybrid scenario generation approach.}
\label{tab:features_scenario_generation}
\begin{tabular}{@{}p{0.35\linewidth} p{0.6\linewidth}@{}}
\toprule
\textbf{Feature} & \textbf{Advantage} \\
\midrule
Forecast anchoring & Captures the underlying temporal structure. \\
Stochastic perturbation & Reflects realistic short-term forecast uncertainty. \\
Clustering & Compresses the scenario space while preserving variability. \\
Probability assignment & Enables risk-weighted cost evaluation. \\
\bottomrule
\end{tabular}
\end{table}

\paragraph{Theoretical Justification.}

This scenario generation methodology builds upon well-established principles in scenario-based robust optimization~\cite{calafiore2006scenario, campi2011exact, bertsimas2011theory} and portfolio-based risk-return trade-offs~\cite{pagnoncelli2009risk}. It has been successfully applied in various energy systems contexts, including microgrids and distributed energy resources~\cite{khodayar2015scenario}. By clustering representative trajectories, the approach strikes a balance between statistical richness and computational tractability—two key requirements for efficient robust MILP formulation.

\subsection{Deterministic Two-Stage MILP Framework (Reference Model)}

The deterministic two-stage MILP framework developed and validated in previous studies~\cite{rojasinnocenti2024electrical, rojas_comparative} optimizes production scheduling and flexibility activation assuming perfect foresight.

\subsubsection{Stage 1: Baseline Production Scheduling}

\paragraph{Problem Structure}\mbox{}

\noindent The system includes:

\begin{itemize}
    \item Flexible production units $k \in \mathcal{K}$,
    \item Product storage silos $i \in \mathcal{S}$,
    \item On-site photovoltaic generation,
    \item Battery Energy Storage System (BESS),
    \item Connection to the public electricity grid.
\end{itemize}

\paragraph{Decision Variables}

\begin{itemize}
    \item $Y_{k,t} \in \{0,1\}$: Production unit on/off status,
    \item $I_{i,t} \geq 0$: Silo inventory level,
    \item $P_{b,t}, P_{s,t} \geq 0$: Grid import/export,
    \item $P_{C,t}, P_{D,t} \geq 0$: BESS charging/discharging powers,
    \item $\mathrm{SOC}_t \geq 0$: BESS state of charge.
\end{itemize}

\paragraph{Objective Function.}

Minimize the total electricity procurement cost over the time horizon:

\begin{multline}
\min\; \Phi = \sum_{t \in \mathcal{T}} \bigg( 
\pi_{b,t} P_{b,t} 
+ \pi_U (P_{C,t} + P_{D,t}) \\
+ \sum_{i \in \mathcal{S}} \pi_{S,i,t} I_{i,t} 
\bigg) \Delta t
\label{eq:obj}
\end{multline}

\paragraph{System Constraints.}

The optimization is subject to the following operational constraints:

\vspace{0.5em}

\noindent\textbf{Mass balance:}
\begin{equation}
\sum_{k \in \mathcal{K}} \Pi_k Y_{k,t} 
+ \sum_{i \in \mathcal{S}} I_{i,t-1} 
= \sum_{i \in \mathcal{S}} I_{i,t},
\quad \forall t \in \mathcal{T}
\label{eq:C1}
\end{equation}

\noindent\textbf{Power balance:}
\begin{equation}
P_{b,t} + P_{D,t} + P_{\mathrm{PV},t} 
= P_{s,t} + P_{C,t} + \sum_{k \in \mathcal{K}} P_k Y_{k,t},
\quad \forall t \in \mathcal{T}
\label{eq:C2}
\end{equation}

\noindent\textbf{Inventory limits:}
\begin{equation}
I_{i,\min} \leq I_{i,t} \leq I_{i,\max},
\quad \forall i \in \mathcal{S},\; t \in \mathcal{T}
\label{eq:C3}
\end{equation}

\noindent\textbf{Demand coverage:}
\begin{equation}
\sum_{i \in \mathcal{S}} I_{i,t} \geq D_t,
\quad \forall t \in \mathcal{T}
\label{eq:C4}
\end{equation}

\noindent\textbf{BESS state-of-charge dynamics:}
\begin{equation}
\mathrm{SOC}_t = \mathrm{SOC}_{t-1} + (P_{C,t} - P_{D,t}) \Delta t,
\quad \forall t \in \mathcal{T}
\label{eq:C5}
\end{equation}

\noindent\textbf{BESS capacity limits:}
\begin{equation}
C_{\max}(1 - \mathrm{DoD}) \leq \mathrm{SOC}_t \leq C_{\max},
\quad \forall t \in \mathcal{T}
\label{eq:C6}
\end{equation}

\noindent\textbf{Charging/discharging power limits:}
\begin{equation}
0 \leq P_{C,t} \leq P_C^{\max}, 
\quad 0 \leq P_{D,t} \leq P_D^{\max},
\quad \forall t \in \mathcal{T}
\label{eq:C7}
\end{equation}

\noindent\textbf{Grid import limits:}
\begin{equation}
0 \leq P_{b,t} \leq P_b^{\max},
\quad \forall t \in \mathcal{T}
\label{eq:C8}
\end{equation}

\noindent\textbf{Minimum uptime:}
\begin{align}
(Y_{k,t+1} - Y_{k,t}) M_k^{\mathrm{ON}} 
&\leq \sum_{j=1}^{M_k^{\mathrm{ON}}} Y_{k,t+j}, \notag \\
&\forall k \in \mathcal{K},\; t \leq N_T - M_k^{\mathrm{ON}}
\label{eq:C9}
\end{align}

\noindent\textbf{Minimum downtime:}
\begin{align}
\sum_{j=1}^{M_k^{\mathrm{OFF}}} Y_{k,t+j} 
&\leq (1 + Y_{k,t+1} - Y_{k,t}) M_k^{\mathrm{OFF}}, \notag \\
&\forall k \in \mathcal{K},\; t \leq N_T - M_k^{\mathrm{OFF}}
\label{eq:C10}
\end{align}

\subsubsection{Stage 2 – Flexibility Activation via SIDC Participation}

\paragraph{Conceptual overview.}
Flexibility is unlocked by adjusting the baseline plan through continuous intraday trading (SIDC), subject to operational feasibility.

\paragraph{New decision variable.}
Let $P_{m,t}$ denote the power traded in the SIDC at time $t$,  
positive for purchases and negative for sales.

\paragraph{Flexible objective function}
\begin{multline}
\min\; \Phi^{\dagger} = \sum_{t \in \mathcal{T}} \Bigl(
  \pi_{b,t} P_{b,t}
  + \pi_{m,t} P_{m,t}
  + \pi_U (P_{C,t} + P_{D,t}) \\
  + \sum_{i \in \mathcal{S}} \pi_{S,i,t} I_{i,t} \Bigr)\,\Delta t
\label{eq:obj_flex}
\end{multline}

\paragraph{Additional constraints}\mbox{}\\[2pt]

\noindent\textbf{SIDC trading window:}
\begin{align}
P_{m,t} &= 0,
&&\forall\, t < \tau_1 \;\text{or}\; t > \tau_2 \label{eq:F1a} \\[0.25em]
|P_{m,t}| &\le LC1,
&&\forall\, t \in [\tau_1,\tau_2] \label{eq:F1b}
\end{align}

\noindent\textbf{Grid import frozen outside window:}
\[
P_{b,t} = P_{b,t}^{*},
\qquad \forall\, t \le \tau_2
\tag{F2}\label{eq:F2}
\]

\noindent\textbf{Updated power balance:}
\begin{align}
P_{b,t} &+ P_{m,t} + P_{D,t} + P_{\mathrm{PV},t} \notag \\
&= P_{s,t} + P_{C,t} 
+ \sum_{k \in \mathcal{K}} P_k Y_{k,t}, \quad \forall\, t \in \mathcal{T}
\label{eq:F3}
\end{align}

\paragraph{Flexibility revenue}
\begin{equation}
\Delta \Phi = \Phi^{*} - \Phi^{\dagger}
\tag{F4}\label{eq:F4}
\end{equation}

A positive value of $\Delta \Phi$ indicates that SIDC participation was profitable.

\subsection{Scenario-Based Robust Two-Stage MILP Framework (New Contribution)}

Building on the deterministic structure, a scenario-based robust MILP model is developed to account for uncertainty in day-ahead electricity prices, PV generation, and production demand.

\paragraph{Decision Structure}

\begin{itemize}[left=0pt]
    \item \textbf{Non-adaptive decisions}: Binary production schedules $Y_{k,t}$ shared across all scenarios.
    \item \textbf{Adaptive decisions}: Operational variables ($P_{b,t}^s$, $P_{C,t}^s$, $P_{D,t}^s$, $I_{i,t}^s$) adapted to each scenario $s \in S$.
\end{itemize}

\subsubsection{First-Stage Robust Baseline Scheduling}

\paragraph{Robust Objective Function}

\begin{equation}
\min \left[ (1-\alpha) \max_{s\in S} \Phi^{\mathrm{BL},s} + \alpha \sum_{s\in S} \rho_s \Phi^{\mathrm{BL},s} \right]
\end{equation}

where $\Phi^{\mathrm{BL},s}$ is the baseline cost in scenario $s$, $\rho_s$ is the probability weight of scenario $s$, and $\alpha \in [0,1]$ controls the risk aversion.

\subsubsection{Second-Stage Robust Flexibility Activation}
\label{sec:2nd_stage_robust}

\paragraph{Robust Flexible Objective Function}

\begin{equation}
\min \left[ (1-\alpha) \max_{s\in S} \Phi^{\mathrm{FLX},s} + \alpha \sum_{s\in S} \rho_s \Phi^{\mathrm{FLX},s} \right]
\end{equation}

where $\Phi^{\mathrm{FLX},s}$ denotes the total flexible production cost under scenario $s$.

\paragraph{Constraints}

All system constraints (mass balance, power balance, BESS dynamics, demand satisfaction, trading limits) are enforced independently for each $(t,s)$ pair, summarized in table~\ref{tab:constraint_comparison}.

\vspace{0.2cm}

\begin{table*}[h]
\centering
\caption{Comparison of constraint application in deterministic vs robust models.}
\label{tab:constraint_comparison}
\begin{tabular}{@{}p{0.28\textwidth} p{0.33\textwidth} p{0.33\textwidth}@{}}
\toprule
\textbf{Constraint type} & \textbf{Deterministic model} & \textbf{Robust model} \\
\midrule
Power balance            & For each $t$ & For each $(t,s)$ (adaptive) \\
BESS dynamics            & For each $t$ & For each $(t,s)$ (adaptive) \\
Inventory dynamics       & For each $t$ & For each $(t,s)$ (adaptive) \\
Demand satisfaction      & For each $t$ & For each $(t,s)$ (adaptive) \\
SIDC trading             & Only in Stage 2 & In Stage 2 across $(t,s)$ (adaptive) \\
Minimum uptime/downtime & For each $t$ & Shared across scenarios (non-adaptive) \\
\bottomrule
\end{tabular}
\end{table*}

\paragraph{Rationale for the Weighted Robust Objective}

Instead of pursuing a purely conservative (min-max) optimization, the proposed framework adopts a weighted robust objective that allows explicit tuning between worst-case protection and expected cost minimization~\cite{ben2009robust, bertsimas2011theory, shapiro2009lectures}. This is achieved through a convex combination governed by the parameter $\alpha$:

\begin{itemize}
    \item $\alpha = 0$: Fully conservative optimization that minimizes the worst-case cost across all scenarios.
    \item $\alpha = 1$: Risk-neutral optimization that minimizes the expected cost, assuming probabilistic weights.
    \item $0 < \alpha < 1$: Risk-averse optimization balancing robustness and performance.
\end{itemize}

This formulation enables the decision-maker to flexibly tailor the optimization strategy based on the system’s criticality, tolerance to risk, and operational context.

\subsection*{Summary of the Methodological Structure}

The framework distinguishes between decision adaptiveness and sequential scheduling stages:

\begin{itemize}[left=0pt]
    \item \textbf{Non-adaptive decisions:}  
    Shared across all scenarios (e.g., unit commitment $Y_{k,t}$); determined before uncertainty is realized.

    \item \textbf{Adaptive decisions:}  
    Scenario-specific variables (e.g., grid imports, BESS dispatch, inventories) that respond to realized conditions.

    \item \textbf{Stage~1 – Robust Baseline Scheduling:}  
    A full robust MILP, where SIDC trading is disabled, but all other decisions—including adaptive ones—are optimized under uncertainty.

    \item \textbf{Stage~2 – Flexibility Activation:}  
    A second robust MILP is solved where early baseline decisions are fixed. SIDC trading is enabled within its window, while the remaining controls are re-optimized per scenario over the rest of the horizon.
\end{itemize}

This framework combines scenario-based uncertainty modeling with a layered optimization approach that distinguishes between non-adaptive and adaptive decisions. By integrating robust scheduling with tunable risk aversion, it ensures feasibility across all scenarios while enabling real-time use of flexibility assets. The two-stage structure delivers reliable non-adaptive planning and adaptive responsiveness—achieving resilience to forecast deviations and capturing value from arbitrage opportunities. This design is particularly suited to industrial systems, where rigid operational constraints coexist with strategically deployable flexibility.

\subsection{Implementation Workflow}

The methodological pipeline consists of the following sequential steps:

\begin{enumerate}[left=0pt]
    \item \textbf{Data Preparation:}  
    Collect historical time series for day-ahead electricity prices, PV generation, and production demand.

    \item \textbf{Forecasting:}  
    Fit seasonal ARIMA models to generate deterministic forecasts of each uncertain variable.

    \item \textbf{Stochastic Simulation:}  
    Generate a large ensemble of trajectories by injecting resampled residuals into the ARIMA forecast.

    \item \textbf{Scenario Reduction:}  
    Normalize, cluster, and rescale the trajectories; assign probabilities based on cluster membership to form a representative scenario set.

    \item \textbf{Robust Optimization:}  
    Solve two sequential robust MILPs:
    \begin{itemize}
        \item \textbf{Stage~1 – Baseline Scheduling:}  
        Optimize the Robust Baseline Scheduling, that is without SIDC participation.        
        \item \textbf{Stage~2 – Flexibility Activation:}  
        Re-optimize under the same scenario set, with SIDC trading enabled and early baseline decisions fixed from Stage~1.
    \end{itemize}
\end{enumerate}

\subsection*{Discussion}

The proposed framework offers a comprehensive and tractable approach for robust industrial energy scheduling under uncertainty. It presents several advantages:

\begin{itemize}[left=0pt]
    \item \textbf{Explicit uncertainty modeling:}  
    Incorporates price, generation, and demand uncertainty through scenario-based representation.

    \item \textbf{Decoupled yet coherent stages:}  
    Maintains consistency between baseline scheduling and flexibility activation through a two-stage formulation.

    \item \textbf{Risk customization:}  
    Allows fine-tuning of the robustness performance tradeoff via the parameter $\alpha$.

\end{itemize}

However, some limitations must be noted. Computational burden increases with the number of scenarios, potentially affecting scalability. Moreover, solution quality is inherently linked to the statistical representativeness of the scenario set—a critical factor when applying the method in real-world settings.

\section{Case Study: Robust Scheduling in a Cement Manufacturing Plant}
\label{sec:case_study}

This section applies the proposed robust optimization framework to a real-world industrial setting, demonstrating its practical relevance and effectiveness under uncertainty. A medium-scale Portland cement manufacturing plant located in Spain serves as the testbed. The focus is placed on the plant’s raw milling subprocess and associated flexibility assets, with the objective of quantifying the operational and economic value of scenario-based robust scheduling in the presence of uncertain electricity prices.

\subsection{Industrial System and Operational Scope}

The selected facility represents a typical energy-intensive cement production plant. Among its processes, the raw milling unit is identified as the most flexible and power-demanding component, making it well-suited for demand-side optimization~\cite{rojas_comparative,rojasinnocenti2024electrical}.

The raw mill consumes approximately \SI{6}{MW} during operation and produces raw meal at a rate of \SI{360}{\tonne\per\hour}. The output is stored in a silo with a total capacity of \SI{15000}{\tonne}, with a minimum inventory requirement of \SI{9000}{\tonne} to ensure uninterrupted feeding of the kiln. The kiln’s demand is a constant inflow \SI{240}{\tonne\per\hour}, reflecting a deterministic operational requirement of the plant.

To enhance flexibility, the plant is equipped with a lithium-ion Battery Energy Storage System of \SI{1}{MWh} capacity, operating at a charge/discharge C-rate of 0.5~h$^{-1}$. While the modeling framework can accommodate uncertainty in on-site photovoltaic generation, PV is excluded in this study to isolate the roles of production and storage flexibility. Key system parameters are summarized in Table~\ref{tab:plant_parameters}.

\begin{table*}[ht]
\centering
\caption{Main operational parameters of the cement plant.}
\label{tab:plant_parameters}
\begin{tabularx}{\textwidth}{@{}>{\raggedright\arraybackslash}p{0.22\textwidth} >{\raggedright\arraybackslash}X >{\centering\arraybackslash}m{0.15\textwidth} >{\centering\arraybackslash}m{0.1\textwidth}@{}}
\toprule
\textbf{Parameter} & \textbf{Description} & \textbf{Value} & \textbf{Units} \\
\midrule
\multicolumn{4}{c}{\textbf{Production system}} \\
\midrule
$P_{\text{mill}}$   & Raw mill power consumption       & 6       & MW \\
$\Pi_{\text{mill}}$ & Raw meal production rate         & 360     & t/h \\
$D_t$               & Constant kiln demand             & 240     & t/h \\
$I_{\max}$          & Maximum silo storage             & 15{,}000  & t \\
$I_{\min}$          & Minimum inventory requirement    & 9{,}000   & t \\
$M^{\mathrm{ON}}$   & Minimum continuous operation     & 6       & h \\
$M^{\mathrm{OFF}}$  & Minimum downtime before restart  & 3       & h \\
\midrule
\multicolumn{4}{c}{\textbf{Energy storage system}} \\
\midrule
BESS Capacity       & Battery storage capacity         & 1       & MWh \\
BESS C-rate         & Charge/discharge rate            & 0.5     & h$^{-1}$ \\
PV Capacity         & Installed PV capacity            & 0       & MW \\
\bottomrule
\end{tabularx}
\end{table*}

\subsection{Model Inputs and Uncertainty Characterization}

The simulation model integrates realistic inputs from both electricity markets and plant operations:

\begin{itemize}[left=0pt]
    \item \textbf{Day-ahead electricity prices:} Hourly price data from the Iberian market operator (OMIE), covering January to April 2025, is used as the sole source of uncertainty. This reflects real-world volatility in short-term energy procurement.

    \item \textbf{SIDC prices:} Prices from the intraday continuous market are treated as deterministic, reflecting their real time visibility up to gate closure in the Spanish bidding zone. We use the transaction level price data published by OMIE on April 28, 2025 to emulate the latest market signals available to plant operators.
    
    \item \textbf{Technical and operational parameters:} All model parameters—including equipment specifications, demand profiles, inventory constraints, and on/off rules—are sourced from validated industrial datasets provided by project collaborators.
\end{itemize}

To explicitly model uncertainty in day-ahead prices, the hybrid scenario generation approach from Section~\ref{sec:scenario_generation} is implemented:

\begin{enumerate}[left=0pt]
    \item \textbf{Forecasting via ARIMA:} Seasonal ARIMA models (\texttt{pmdarima} in Python) are used to capture hourly seasonality and trend dynamics in the historical series. We set $m = 24$ to represent the seasonal period (24 hours), and apply $D = 1$ for seasonal differencing. Model orders are selected automatically based on the Akaike Information Criterion (AIC), which balances model fit and complexity to prevent overfitting.

    \item \textbf{Stochastic Perturbation:}  
    Forecast residuals are randomly sampled and injected into the ARIMA forecasts to produce 500 stochastic price trajectories. An inflation factor $\beta = 1.5$ is used to emphasize extreme deviations, in line with robust optimization principles~\cite{calafiore2006scenario, campi2011exact, khodayar2015scenario, zhang2013review}.

    \item \textbf{Clustering and Scenario Reduction:}  
    The ensemble is normalized to remove scale bias and clustered using K-Means (\texttt{sklearn.cluster} in Python). The number of clusters is selected via the Elbow method and Silhouette score analysis. Each centroid forms a scenario, with its probability derived from cluster membership frequency.
\end{enumerate}

This method produces a compact, representative scenario set that supports tractable optimization while preserving market variability.

\subsection{Simulation Framework and Execution}

To reflect practical industrial workflows, a two-stage fixed-horizon simulation structure is adopted:

\begin{itemize}[left=0pt]
    \item \textbf{Stage 1 (Baseline Scheduling):} At the beginning of each simulation cycle, a 7-day robust production schedule is computed, serving as the baseline commitment plan.  
    \item \textbf{Stage 2 (Flexibility Activation):} The plan is refined by activating flexibility, with SIDC trading restricted to a 24-hour window, while BESS and other assets operate over the full horizon.
\end{itemize}

This hierarchical structure reflects real-world planning workflows, in which strategic commitments are made in advance, while operational recourse is exploited in near real-time.

To investigate the effect of risk preference, simulations are conducted across a range of values for the robustness parameter $\alpha \in [0,1]$, enabling a detailed trade-off analysis.

All optimization problems are solved using the \textsc{SCIP} solver~\cite{SCIP_solver, SCIP_solver_Python}, accessed via \texttt{PySCIPOpt}. Simulations were executed on a workstation equipped with an Intel\textsuperscript{\textregistered} Core\texttrademark{} i5-1135G7 CPU (2.4–4.2~GHz) and 16~GB of RAM, running Windows~11 Pro. This configuration provides context for the reported solution times and confirms the tractability of the model on standard computing hardware.

\subsection{Experimental Scope and Assumptions}

The scope of the case study is defined by a set of focused assumptions that ensure analytical clarity while preserving practical relevance. The analysis centers on the cement industry and uses real operational data from a single facility. Although the modeling framework is generalizable, this industrial focus enables a high fidelity evaluation of robust scheduling in a realistic setting.

Only two flexible assets are activated in the simulations: the raw mill and the BESS. PV generation is deliberately excluded to isolate the specific contributions of storage and process level flexibility. Regarding uncertainty, only day-ahead electricity prices are modeled as stochastic variables. All other inputs—including internal demand, technical specifications, and operational constraints—are treated as deterministic.

SIDC prices are assumed to be known within the flexibility window, consistent with their real time visibility and the market design that allows operators to react up to gate closure. All scheduling and trading decisions strictly adhere to system level feasibility requirements, including ramp limits, energy balances, storage dynamics, and SIDC trading rules.

These assumptions narrow the analytical focus, allowing for precise evaluation of how robust optimization affects scheduling outcomes under market price uncertainty. Future work may relax some of these simplifications to capture broader system dynamics and interactions across assets and sites.

The next section presents and discusses the simulation results, highlighting the economic and operational implications of different robustness levels in the energy scheduling process.

\section{Results and Discussion}
\label{sec:results}

This section presents the main findings from the simulation experiments. The analysis is structured in three parts. First, we examine the generated electricity price scenarios. Second, we evaluate the cost-performance trade-off across different risk aversion levels via an $\alpha$-sweep. Finally, we assess the operational behavior of flexibility assets—including the BESS, the raw mill, and SIDC market participation—under selected risk preferences.

\subsection{Scenario Analysis}
\label{subsec:scen_analys}

To model uncertainty in day-ahead electricity prices, we implement a hybrid scenario generation procedure that combines statistical forecasting with stochastic simulation and data-driven reduction. This enables the construction of a compact yet expressive set of price trajectories for robust optimization.

\paragraph{Synthetic Ensemble Generation.}  
The process begins with a seasonal ARIMA model trained on historical market prices, producing a deterministic forecast that captures autocorrelation, seasonality, and long-term trends. Forecast residuals are then sampled and injected into the predicted series, inflated by a factor $\beta = 1.5$ to account for market volatility. This results in an ensemble of 500 stochastic price trajectories, each spanning a 7-day horizon (168 hourly points), reflecting realistic short-term price dynamics.

\begin{figure}[ht]
\centering
\includegraphics[width=1\linewidth]{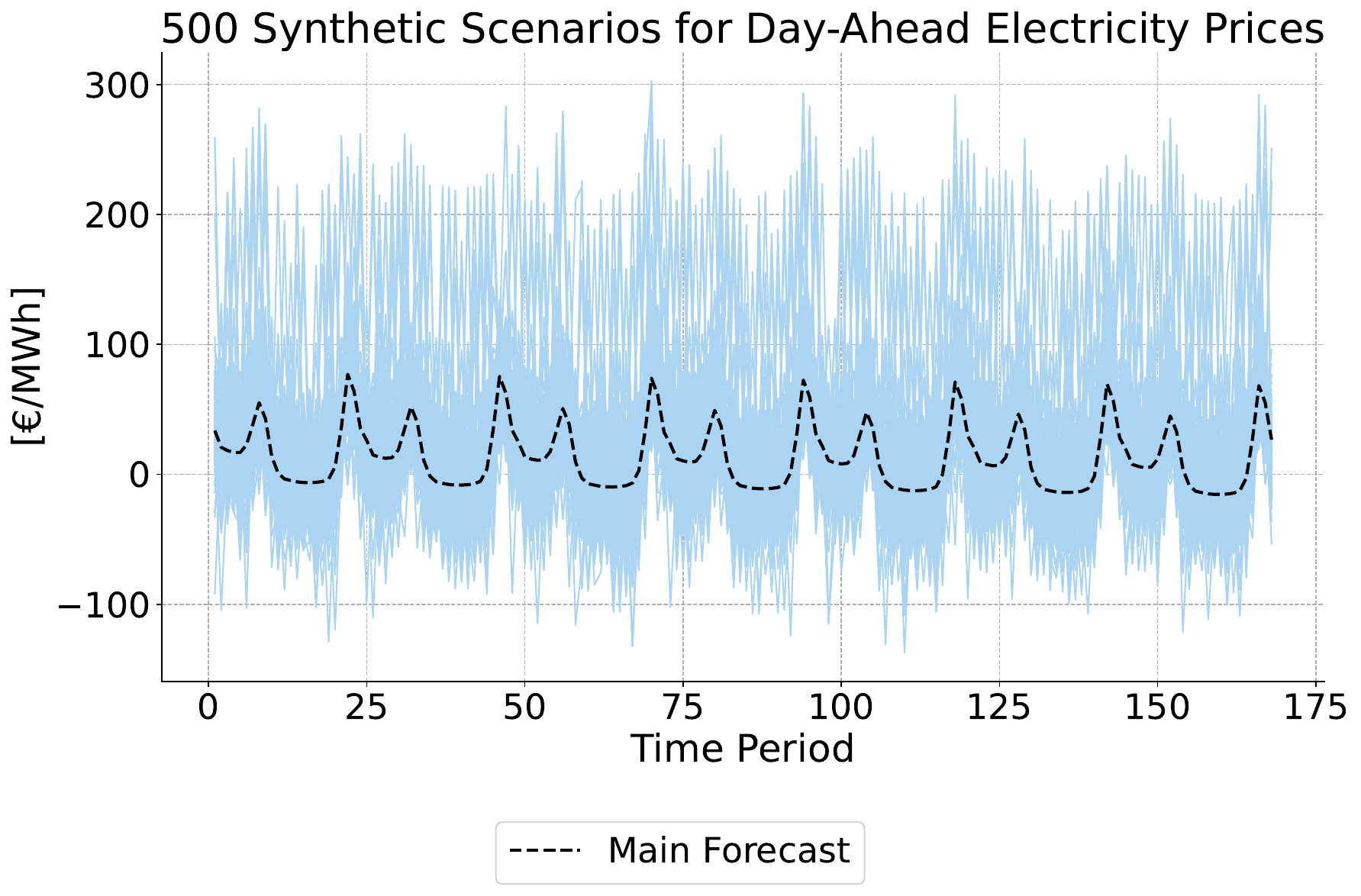}
\includegraphics[width=1\linewidth]{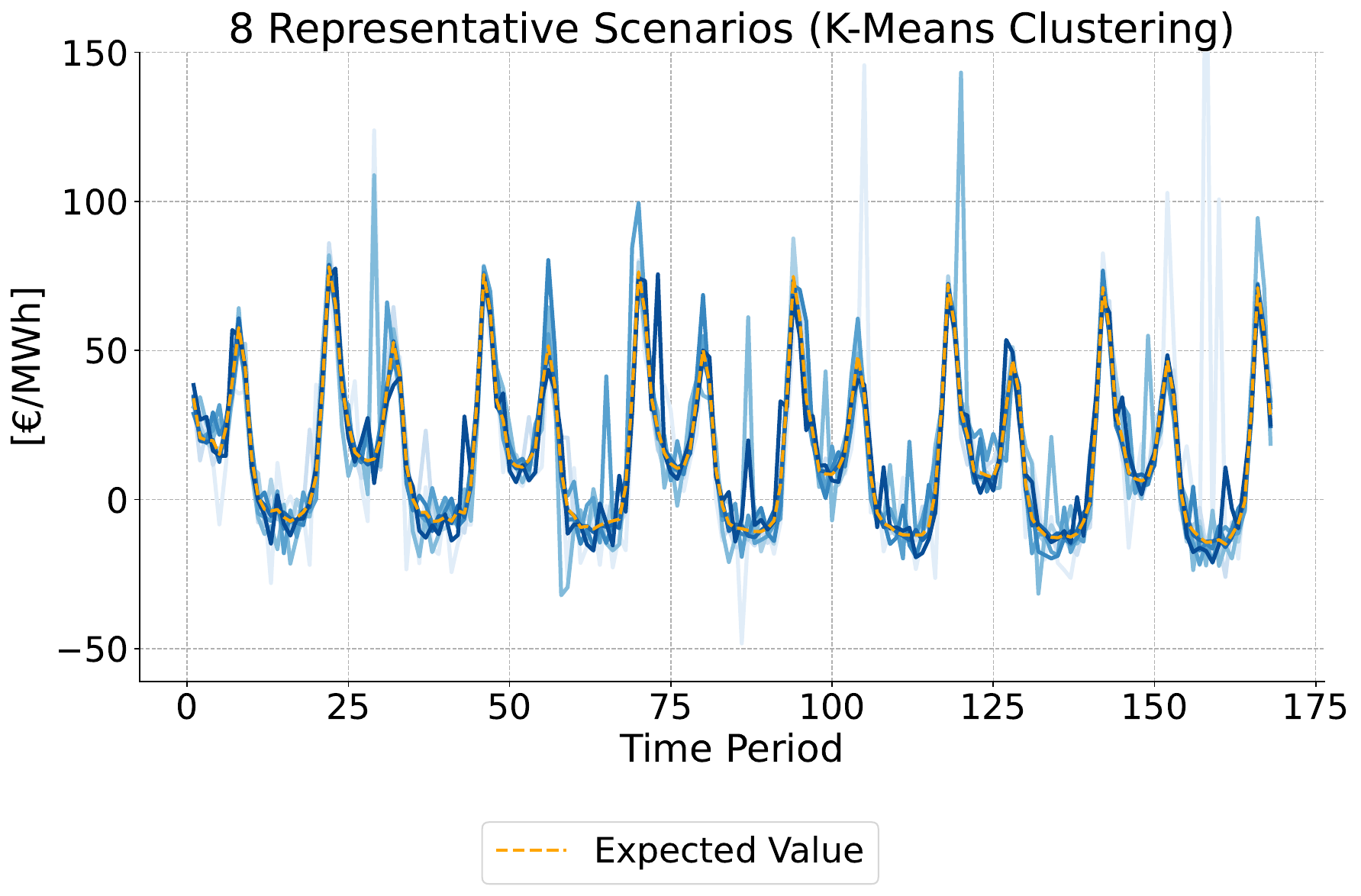}
\caption{Top: Full ensemble of 500 stochastic price trajectories and baseline ARIMA forecast. Bottom: Reduced set of 8 representative scenarios. The expected price profile is shown in dashed orange. Trajectory opacity reflects scenario probability.}
\label{fig:price_scenarios}
\end{figure}

Figure~\ref{fig:price_scenarios} (top) illustrates this ensemble (light blue), alongside the main forecast (dashed black). The variability across trajectories captures potential market conditions, but their volume is computationally prohibitive for direct inclusion in the MILP formulation.

\paragraph{Scenario Reduction via Clustering.}  
To address tractability, we reduce the ensemble to a representative subset using K-Means clustering. Each cluster centroid defines a scenario, and its probability is proportional to the number of trajectories in the cluster.

\begin{figure}[ht]
\centering
\includegraphics[width=1\linewidth]{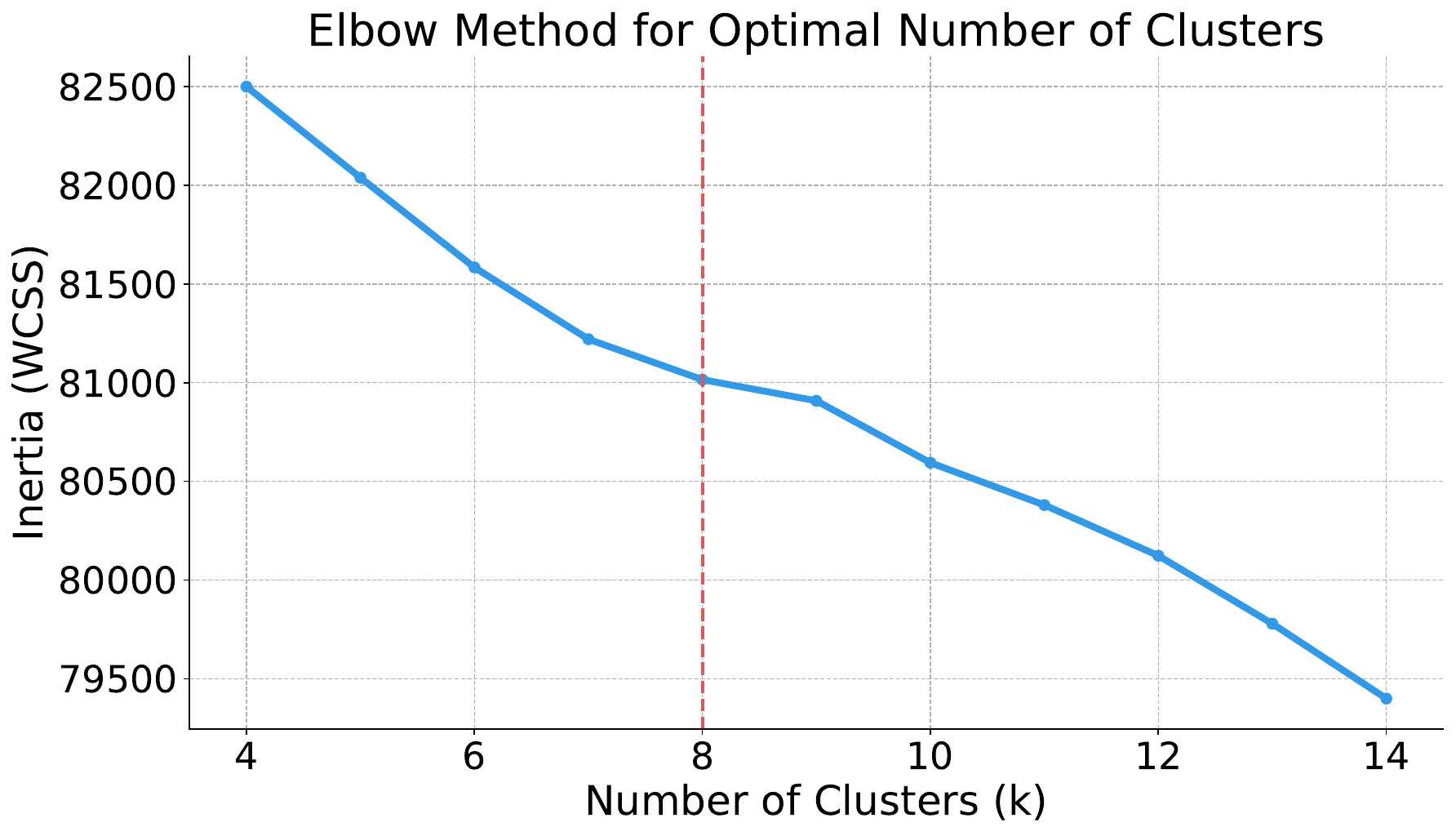}
\includegraphics[width=1\linewidth]{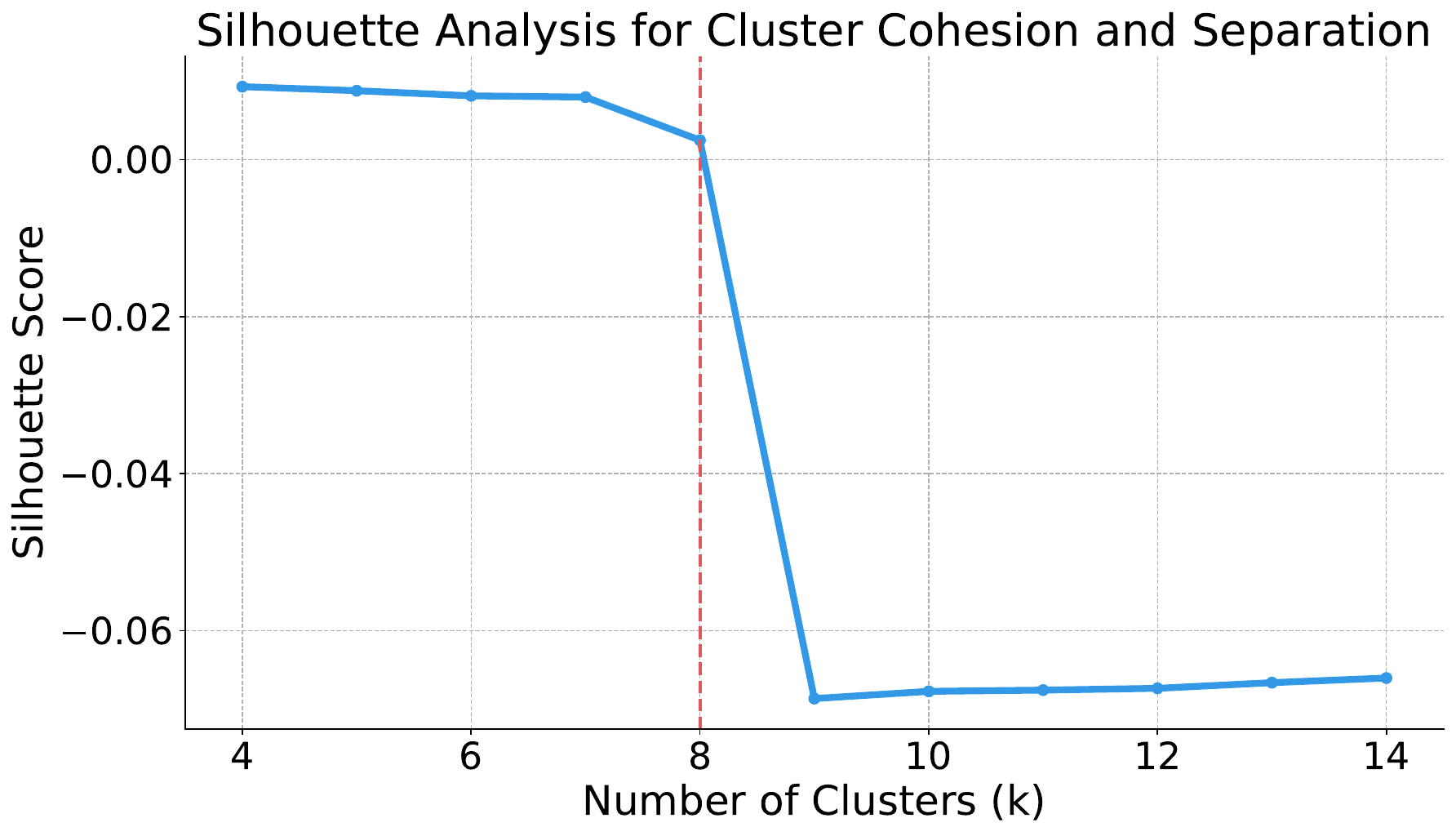}
\caption{Selection of the number of clusters. Top: Elbow method identifies diminishing returns beyond \(k = 8\). Bottom: Silhouette scores indicate reduced clustering quality after \(k = 8\).}
\label{fig:clust_selec}
\end{figure}

The number of clusters \(k\) is determined using two complementary diagnostics: the Elbow method, which identifies the point at which adding more clusters yields diminishing returns in reducing intra-cluster variance (inertia), and the Silhouette score, which evaluates how well each point fits within its assigned cluster. As shown in Figure~\ref{fig:clust_selec}, both methods indicate a turning point at \(k = 8\), suggesting a suitable trade-off between scenario diversity and computational efficiency.

\paragraph{Final Scenario Set.}  
Figure~\ref{fig:price_scenarios} (bottom) presents the reduced scenario set used in the robust MILP. Each trajectory corresponds to a cluster centroid and is assigned a weight based on its frequency. The expected price profile—obtained as the weighted average—is shown in dashed orange. Opacity encodes probability: darker trajectories are more likely.

This reduced set preserves key statistical features of the ensemble, including peak hours, daily fluctuations, and low-probability extremes, enabling reliable and efficient optimization.

\paragraph{Scenario Probabilities.}  
Table~\ref{tab:representative_scenarios} summarizes the scenario probabilities, ordered by frequency. Scenario~6 dominates the distribution (61.6\%), followed by a few moderately likely trajectories. Scenario~4, highlighted in the table, represents the price profile with the highest cumulative cost, despite its low probability of occurrence (3.8\%).

\begin{table}[ht]
\centering
\caption{Scenario probabilities after clustering.}
\label{tab:representative_scenarios}
\begin{tabular}{cc}
\toprule
\textbf{Scenario} & \textbf{Probability (\%)} \\
\midrule
s6 & 61.6 \\
s2 & 15.2 \\
s5 & 8.2 \\
s1 & 6.2 \\
\rowcolor{lightgray}
s4 & 3.8 \\
s7 & 3.8 \\
s3 & 0.8 \\
s0 & 0.4 \\
\bottomrule
\end{tabular}
\end{table}

\paragraph{Implications for Optimization.}  
By optimizing over this reduced and weighted scenario set, the robust MILP ensures feasibility under diverse price realizations. The worst-case scenario in terms of total electricity cost does not necessarily align with the highest-price trajectory; rather, it emerges from interactions between prices, asset constraints, and operational decisions.

This structured treatment of uncertainty enables the model to anticipate market risk while preserving computational tractability—an essential feature for deployment in industrial scheduling environments. The next section explores how this scenario structure translates into robust planning outcomes across varying levels of risk aversion.

\subsection{Robustness–Efficiency Trade-off Analysis}
\label{subsec:alpha_sweep}

To evaluate how risk preferences influence scheduling decisions and cost performance, we conduct a sensitivity analysis over the robustness parameter $\alpha \in [0,1]$. As introduced in the methodology, $\alpha$ governs the trade-off between minimizing the expected operational cost ($\alpha = 1$) and protecting against the worst-case scenario cost ($\alpha = 0$). Intermediate values yield risk-averse schedules.

\begin{figure}[ht]
\centering
\includegraphics[width=1\linewidth]{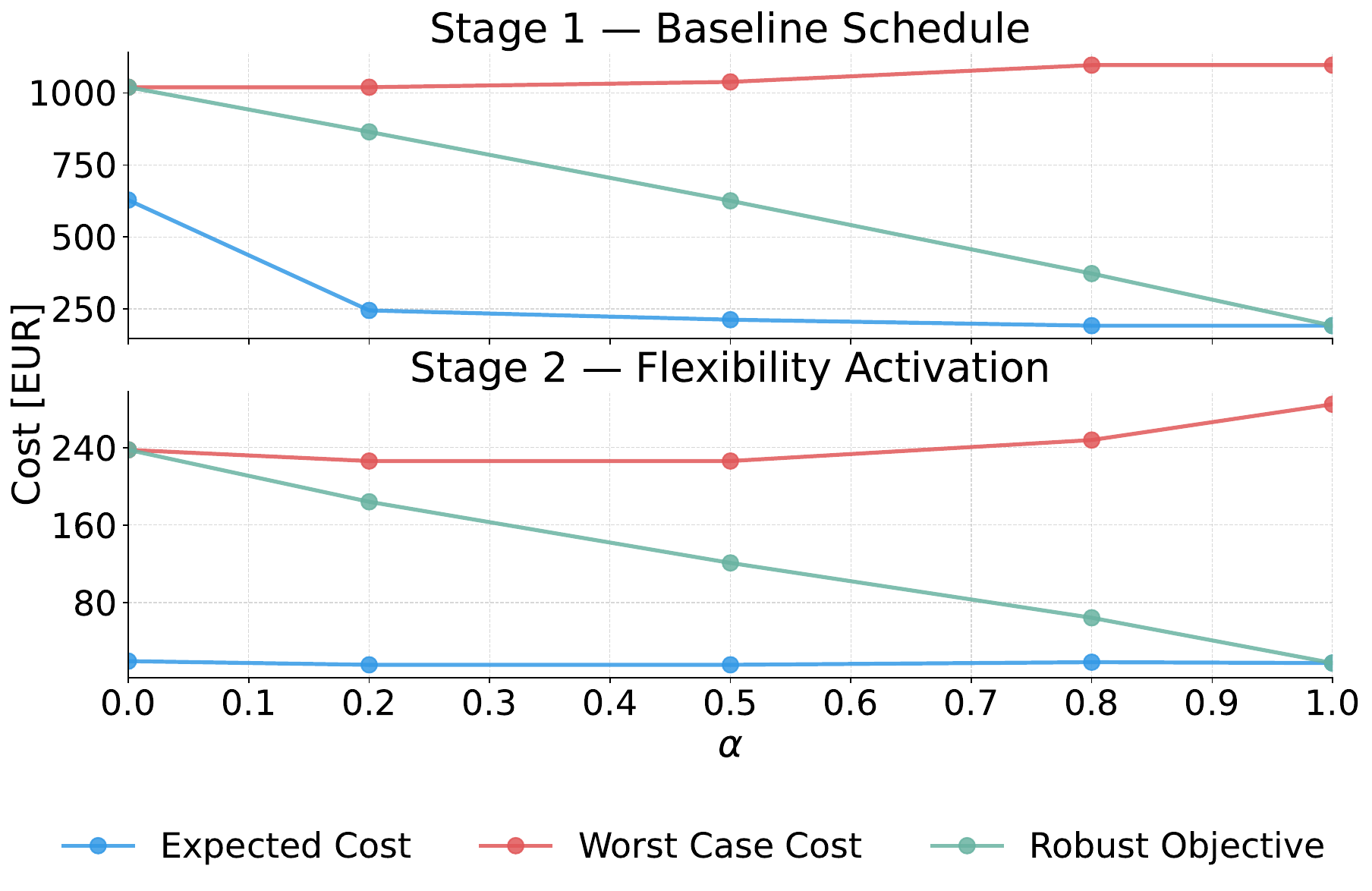}
\caption{Trade-off between expected cost and worst-case cost across values of $\alpha$. Results shown for Stage~1 (baseline scheduling) and Stage~2 (flexibility activation).}
\label{fig:alpha_tradeoff}
\end{figure}

The optimization problem is solved for five representative values of $\alpha$ (0, 0.2, 0.5, 0.8, 1) using the reduced set of 8 representative price scenarios. The results are presented in Figure~\ref{fig:alpha_tradeoff}, showing the evolution of expected cost, worst-case cost, and the resulting robust objective for both scheduling stages.

\paragraph{Key Observations.}

\begin{itemize}[left=0pt]
    \item \textbf{Monotonic behavior.} As $\alpha$ increases, the robust objective decreases in both stages, reflecting the shift from conservative (worst-case-oriented) to risk-neutral optimization.
    
    \item \textbf{Value of flexibility.} For every $\alpha$, the flexible schedule (Stage~2) achieves lower costs than the baseline schedule (Stage~1), highlighting the economic value of SIDC participation and BESS operation.
    
    \item \textbf{Risk sensitivity.} The slope of the cost curves is steeper in Stage~1, meaning baseline scheduling is more sensitive to changes in risk preference. Stage~2 exhibits flatter trends, indicating flexibility assets help buffer cost variability.
    
    \item \textbf{Performance bounds.} At $\alpha = 0$, Stage~1 yields a worst-case cost of \SI{1020.15}{\euro}, while Stage~2 reduces this to \SI{237.69}{\euro}. At $\alpha = 1$, expected costs are as low as \SI{191.84}{\euro} and \SI{17.52}{\euro}, respectively.
\end{itemize}

\paragraph{Numerical Results.}  
Tables~\ref{tab:cost_stage1} and~\ref{tab:cost_stage2} report the exact cost metrics used to compute the curves in Figure~\ref{fig:alpha_tradeoff}.

\begin{table}[ht]
\centering
\caption{Cost metrics for Stage~1 (baseline scheduling) across $\alpha$ values.}
\label{tab:cost_stage1}
\begin{tabularx}{\linewidth}{@{}c >{\centering\arraybackslash}X >{\centering\arraybackslash}X >{\centering\arraybackslash}X@{}}
\toprule
\textbf{$\alpha$} & \textbf{Expected Cost (€)} & \textbf{Worst-Case Cost (€)} & \textbf{Robust Objective (€)} \\
\midrule
0.0 & 628.20 & 1020.15 & 1020.15 \\
0.2 & 244.78 & 1020.15 & 865.08 \\
0.5 & 212.55 & 1038.57 & 625.56 \\
0.8 & 191.84 & 1096.86 & 372.85 \\
1.0 & 191.84 & 1096.86 & 191.84 \\
\bottomrule
\end{tabularx}
\end{table}

\begin{table}[ht]
\centering
\caption{Cost metrics for Stage~2 (flexibility activation) across $\alpha$ values.}
\label{tab:cost_stage2}
\begin{tabularx}{\linewidth}{@{}c >{\centering\arraybackslash}X >{\centering\arraybackslash}X >{\centering\arraybackslash}X@{}}
\toprule
\textbf{$\alpha$} & \textbf{Expected Cost (€)} & \textbf{Worst-Case Cost (€)} & \textbf{Robust Objective (€)} \\
\midrule
0.0 & 19.46 & 237.69 & 237.69 \\
0.2 & 15.67 & 226.15 & 184.05 \\
0.5 & 15.67 & 226.15 & 120.91 \\
0.8 & 18.38 & 247.83 & 64.27 \\
1.0 & 17.52 & 284.78 & 17.52 \\
\bottomrule
\end{tabularx}
\end{table}

\paragraph{Interpretation.}

Moderate values of $\alpha$ (e.g., 0.5–0.8) offer the best trade-offs: they achieve substantial reductions in worst-case cost relative to risk-neutral solutions while avoiding the overly conservative outcomes of min–max strategies. The inclusion of flexibility mechanisms in Stage~2 further enhances this trade-off, delivering both lower expected costs and greater robustness.

This analysis confirms that the convex risk-weighting scheme effectively balances cost-efficiency and protection, making it suitable for industrial settings where economic performance must be safeguarded against market volatility.

\subsubsection{Robust vs. Deterministic Scheduling Strategies}
\label{subsubsec:det_vs_robust}

To further assess the benefits of robust optimization, we compare its performance against a deterministic scheduling strategy across three representative risk-aversion levels: $\alpha \in \{0,\;0.5,\;1\}$. This comparison focuses exclusively on Stage~1 (baseline scheduling).

The \textbf{deterministic strategy} is computed by optimizing the MILP over the expected price profile—that is, the weighted average of all scenario trajectories—thus assuming perfect foresight of average conditions. Once this single schedule is obtained, it is evaluated \emph{ex post} across all scenarios, without re-optimization, to analyze its performance under uncertainty. The comparison includes several performance metrics: expected cost, worst-case cost, cost variability (expressed as weighted standard deviation), battery usage, and solver time.

To ensure that the reported variability captures not only the dispersion of cost outcomes but also their likelihood, the standard deviation is computed using scenario probabilities as weights. This probabilistic formulation offers a more accurate and meaningful measure of uncertainty, in line with the scenario-based nature of the modeling framework.

Importantly, because the deterministic schedule is fixed and does not depend on $\alpha$, its performance metrics remain constant across rows. In contrast, the robust strategy is re-optimized for each $\alpha$ value, adapting decisions to the level of risk aversion. The results are presented in Table~\ref{tab:deterministic_vs_robust}.

\begin{table*}[ht]
\centering
\caption{Performance comparison of deterministic and robust strategies in Stage~1 across selected $\alpha$ values.}
\label{tab:deterministic_vs_robust}
\begin{tabularx}{\textwidth}{@{}c c >{\centering\arraybackslash}X >{\centering\arraybackslash}X >{\centering\arraybackslash}X >{\centering\arraybackslash}m{1.5cm} >{\centering\arraybackslash}m{1.5cm}@{}}
\toprule
\textbf{$\alpha$} & \textbf{Strategy} & \makecell{\textbf{Worst-Case} \\ \textbf{Cost (€)}} & \makecell{\textbf{Expected} \\ \textbf{Cost (€)}} & \makecell{\textbf{Weighted} \\ \textbf{Std. Dev. (€)}} & \makecell{\textbf{Battery} \\ \textbf{Use (MWh)}} & \makecell{\textbf{Solve} \\ \textbf{Time (s)}} \\
\midrule
0   & Deterministic & 1248.04 & 171.62 & 308.11 & 12.80 & 1.40 \\
    & Robust        & 1221.62 & 359.79 & 253.13 & 27.16 & 38.41 \\
\midrule
0.5 & Deterministic & 1248.04 & 171.62 & 308.11 & 12.80 & 1.78 \\
    & Robust        & 1223.23 & 266.76 & 295.55 & 32.40 & 22.67 \\
\midrule
1   & Deterministic & 1248.04 & 171.62 & 308.11 & 12.80 & 1.32 \\
    & Robust        & 1285.30 & 248.30 & 296.85 & 33.20 & 18.35 \\
\bottomrule
\end{tabularx}
\end{table*}

\paragraph{Interpretation.}

The comparative results highlight several meaningful trends:

\begin{itemize}[left=0pt]
    \item \textbf{Robust strategies consistently reduce worst-case costs} when $\alpha < 1$. For example, at $\alpha = 0$, the worst-case cost drops from \SI{1248.04}{\euro} (deterministic) to \SI{1221.62}{\euro} (robust).
    
    \item \textbf{Expected costs for robust strategies are higher}, reflecting a deliberate trade-off between average performance and protection against extreme outcomes. At $\alpha = 0.5$, the robust strategy incurs an expected cost of \SI{266.76}{\euro}, compared to \SI{171.62}{\euro} in the deterministic case.

    \item \textbf{Battery usage nearly triples} under robust optimization, increasing from \SI{12.8}{MWh} to over \SI{33}{MWh} at $\alpha = 1$, suggesting a more aggressive use of storage to buffer uncertainty.

    \item \textbf{Standard deviation behavior varies with risk preference.} At $\alpha = 0$, robust scheduling leads to a lower cost variability than the deterministic case (\SI{253.13}{\euro} vs.\ \SI{308.11}{\euro}), but at $\alpha = 1$, variability slightly increases (\SI{296.85}{\euro}), indicating that risk-neutral strategies may result in higher exposure to volatility.

    \item \textbf{Computation times increase significantly} for robust strategies due to multi scenario enumeration. Nevertheless, all cases remain computationally tractable for real time industrial scheduling.
\end{itemize}

\paragraph{Clarification on Stage~2.}

A direct deterministic comparison is not provided for Stage~2 because this stage builds upon the Stage~1 schedule and introduces SIDC market participation. The baseline in Stage~2 refers to a robust model without intraday trading (see Section~\ref{sec:2nd_stage_robust}), and thus no deterministic counterpart exists. Therefore, performance analysis in Stage~2 focuses exclusively on robust strategies under varying $\alpha$ values.

\vspace{0.2cm}

This comparison reinforces the practical advantage of risk-aware planning: while deterministic models offer simplicity and lower computational effort, they perform poorly in extreme scenarios. Robust optimization, by contrast, delivers more resilient schedules at the cost of moderately higher average expense and solve time—a trade-off that is often acceptable in mission-critical industrial settings.

\subsection{Operational Impact of Risk Aversion Preferences}
\label{subsec:operational_behavior}

This subsection analyzes how key flexible assets—the BESS, the raw mill, and intraday trading via the SIDC—respond to varying degrees of risk aversion, governed by the parameter $\alpha$. For each configuration $\alpha \in \{0.0,\; 0.5,\; 1.0\}$, we examine the corresponding operational behavior under the worst-case scenario associated with each robust strategy. This approach reveals how non-adaptive scheduling decisions shape adaptive asset-level operations under adverse price conditions.

\subsubsection{Battery Operation Patterns}

\begin{figure}[ht]
\centering
\includegraphics[width=1\linewidth]{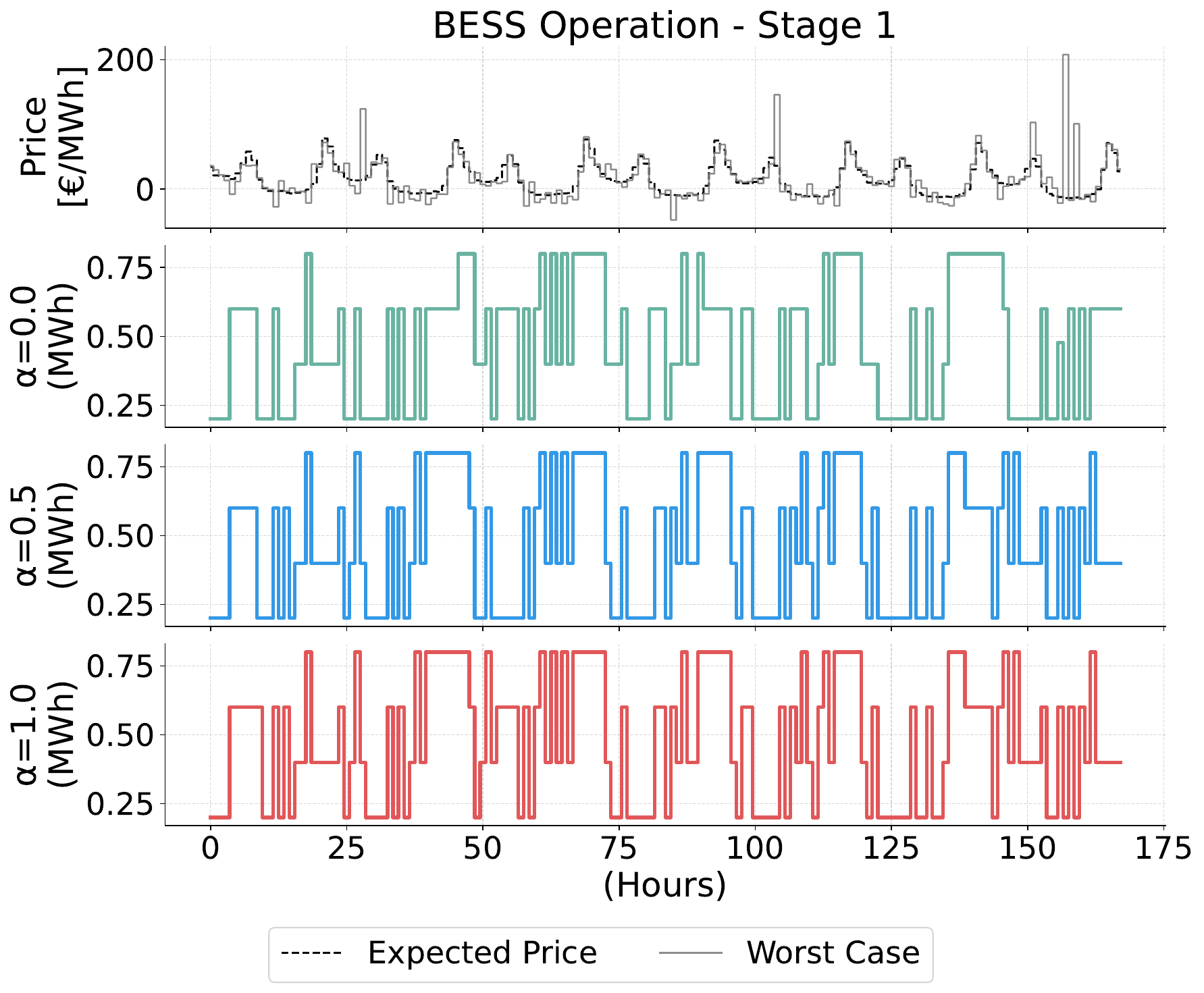}
\includegraphics[width=1\linewidth]{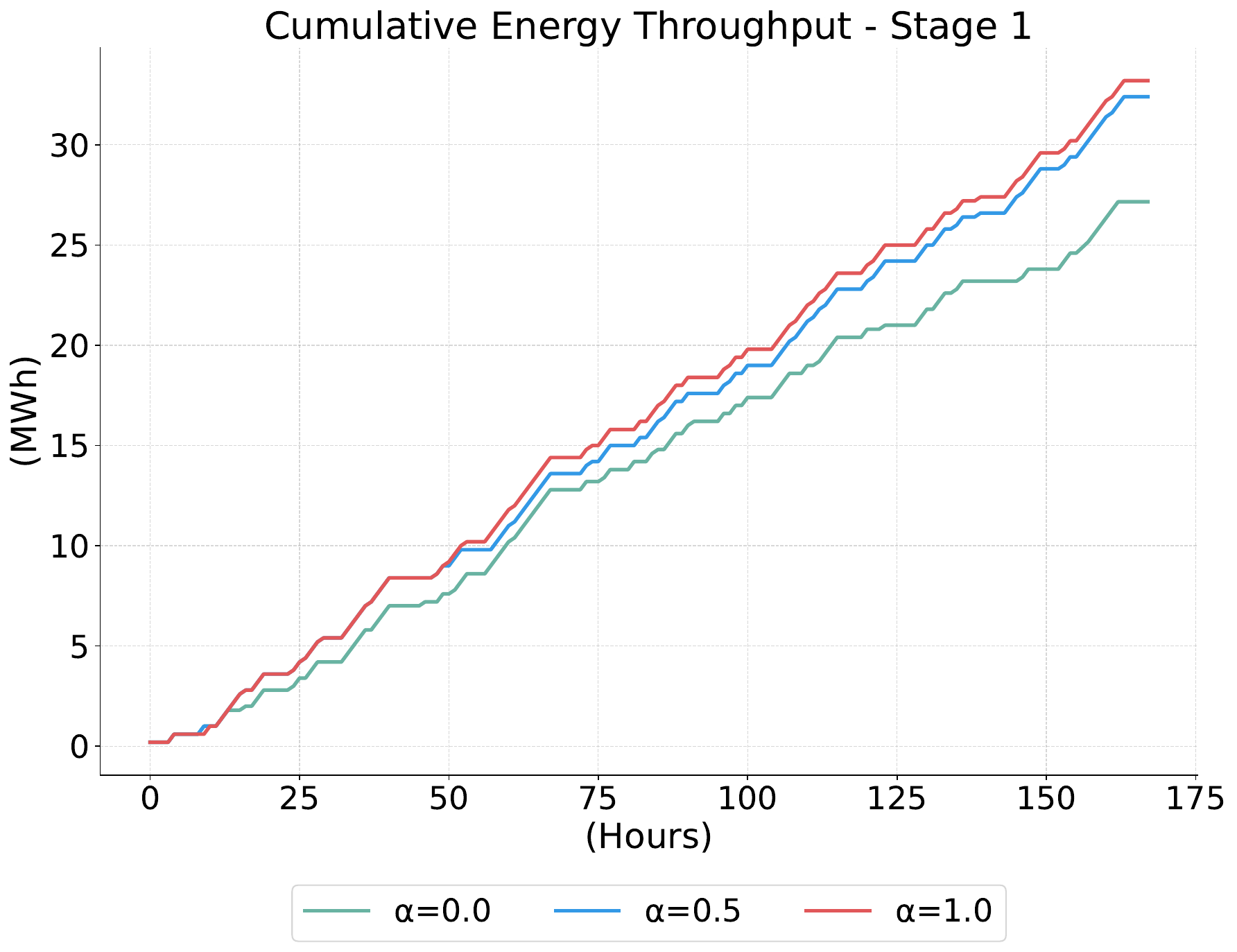}
\caption{Top: Hourly BESS SoC profiles in Stage~1 under worst-case scenario. Bottom: Cumulative battery throughput across $\alpha$ values.}
\label{fig:bess_soc_stage1}
\end{figure}

Figure~\ref{fig:bess_soc_stage1} (top) presents the BESS state-of-charge (SoC) trajectories in Stage~1 (baseline scheduling), under each strategy’s respective worst-case scenario. The top panel displays both expected and worst-case day-ahead price profiles, while the lower panels illustrate the resulting SoC evolution across three risk aversion levels. Although actual BESS operation depends on scenario realization, evaluating all cases under worst-case conditions provides a meaningful comparison of how non-adaptive decisions condition adaptive system behavior when exposed to extreme volatility.

Clear patterns emerge as $\alpha$ varies:

\begin{itemize}[left=0pt]
    \item \textbf{Risk-neutral strategy ($\alpha = 1.0$):} The battery performs frequent and deep cycles, aggressively pursuing arbitrage opportunities in response to price variations.

    \item \textbf{Conservative strategy ($\alpha = 0.0$):} Battery operation is more cautious, exhibiting shallower cycles and reduced throughput to ensure feasibility across all price realizations.
\end{itemize}

These trends are quantified in Figure~\ref{fig:bess_soc_stage1} (bottom), which plots the \textbf{cumulative energy throughput}—the total amount of energy charged and discharged over time. This metric acts as a proxy for battery utilization intensity. Throughput reaches approximately \SI{33}{MWh} in the risk-neutral case, compared to about \SI{27}{MWh} under the fully robust configuration, indicating that risk appetite directly shapes battery deployment.

\begin{figure}[ht]
\centering
\includegraphics[width=1\linewidth]{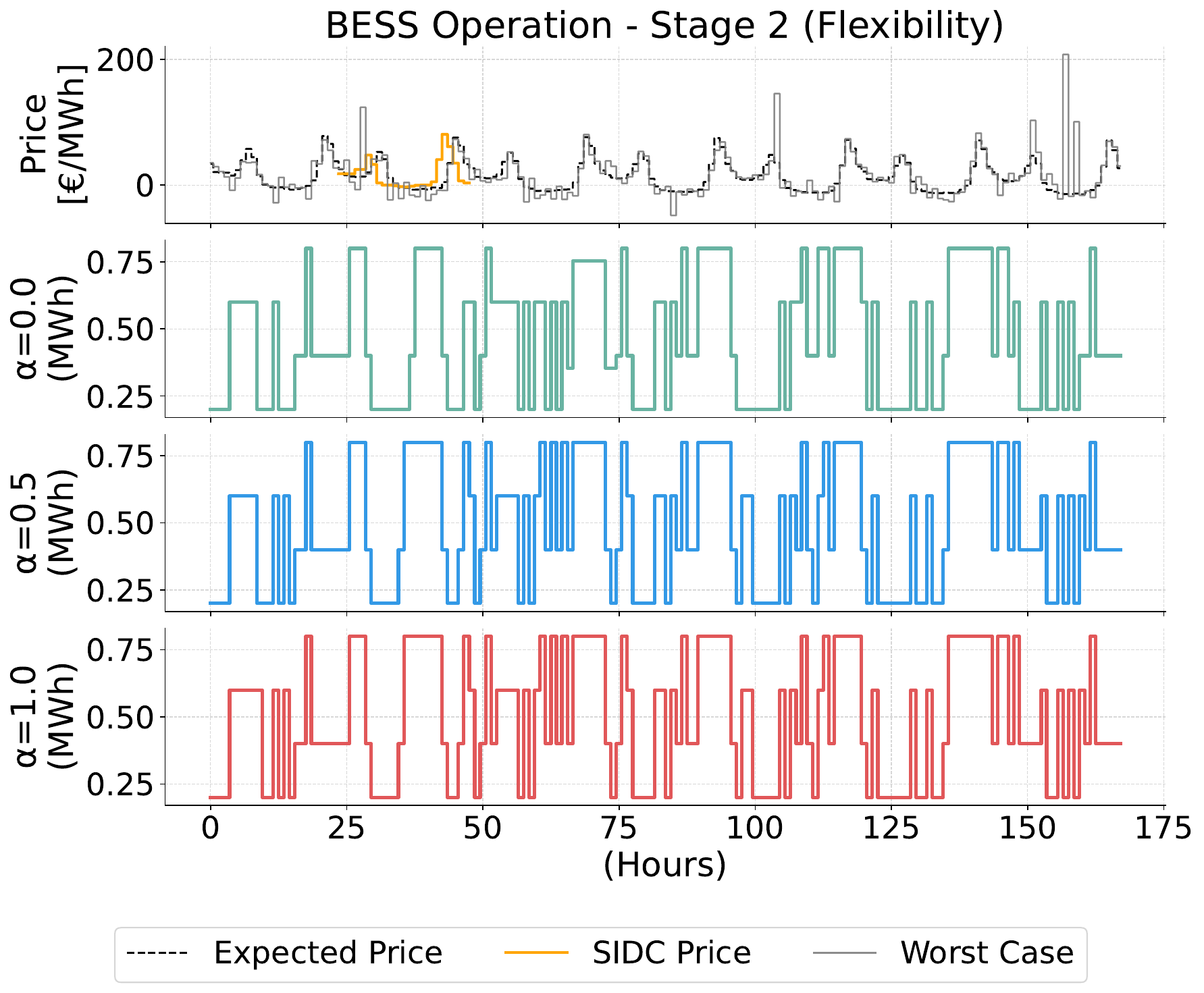}
\includegraphics[width=1\linewidth]{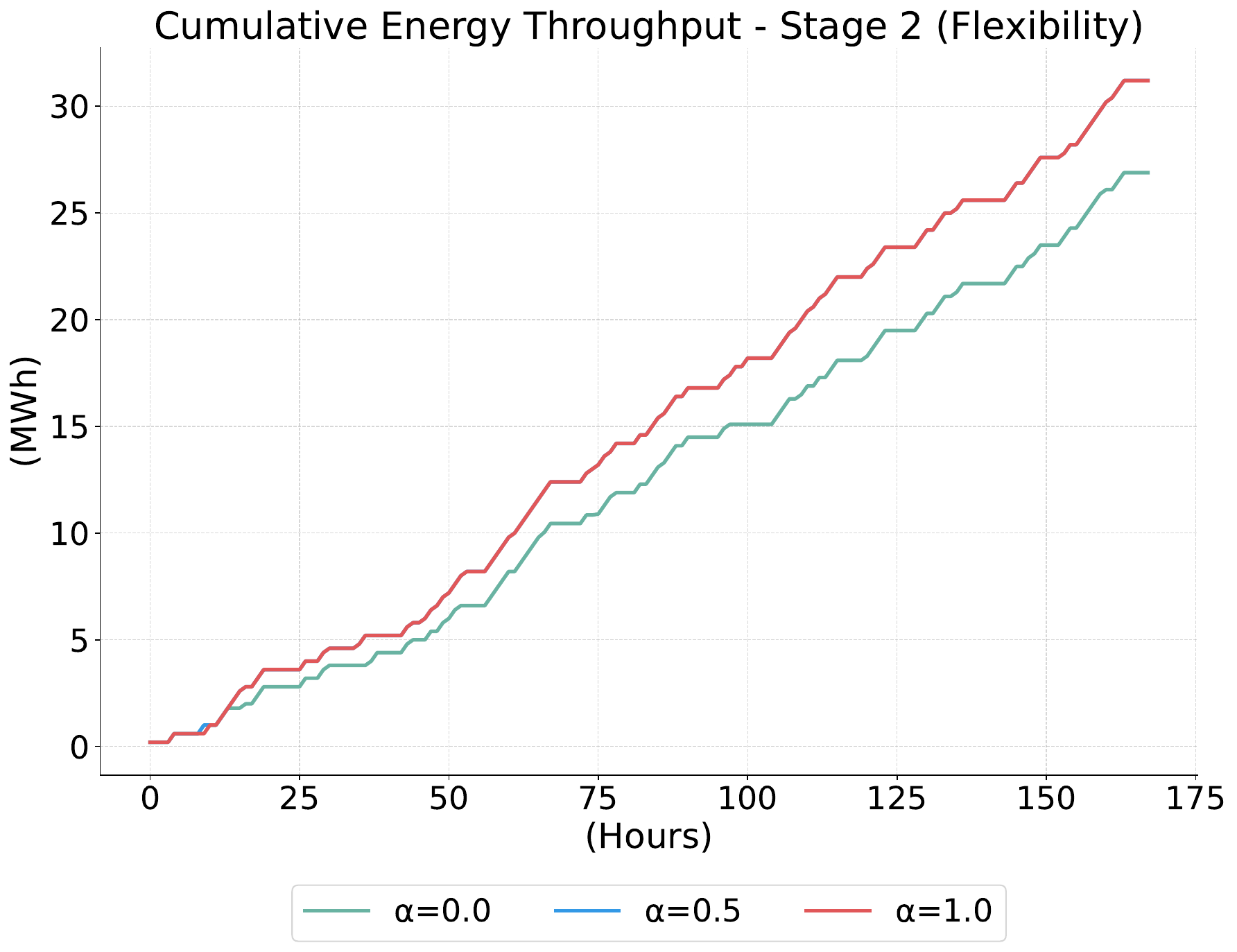}
\caption{Top: Hourly BESS SoC profiles in Stage~2 (with SIDC) under worst-case scenario. Bottom: Cumulative battery throughput across $\alpha$ values.}
\label{fig:bess_soc_stage2}
\end{figure}

Stage~2 results follow a similar pattern (Figure~\ref{fig:bess_soc_stage2}). With SIDC trading allowed in hours 24–48, the system gains additional degrees of freedom for arbitrage. The SoC profiles for $\alpha = 0.5$ and $\alpha = 1.0$ closely align, while the $\alpha = 0.0$ case exhibits restrained engagement. These findings confirm that the BESS dynamically adapts its operation to the risk profile encoded in the scheduling stage, thereby acting as a key instrument for economic optimization under uncertainty.

\subsubsection{SIDC Market Participation Behavior}

\begin{figure}[ht]
\centering
\includegraphics[width=1\linewidth]{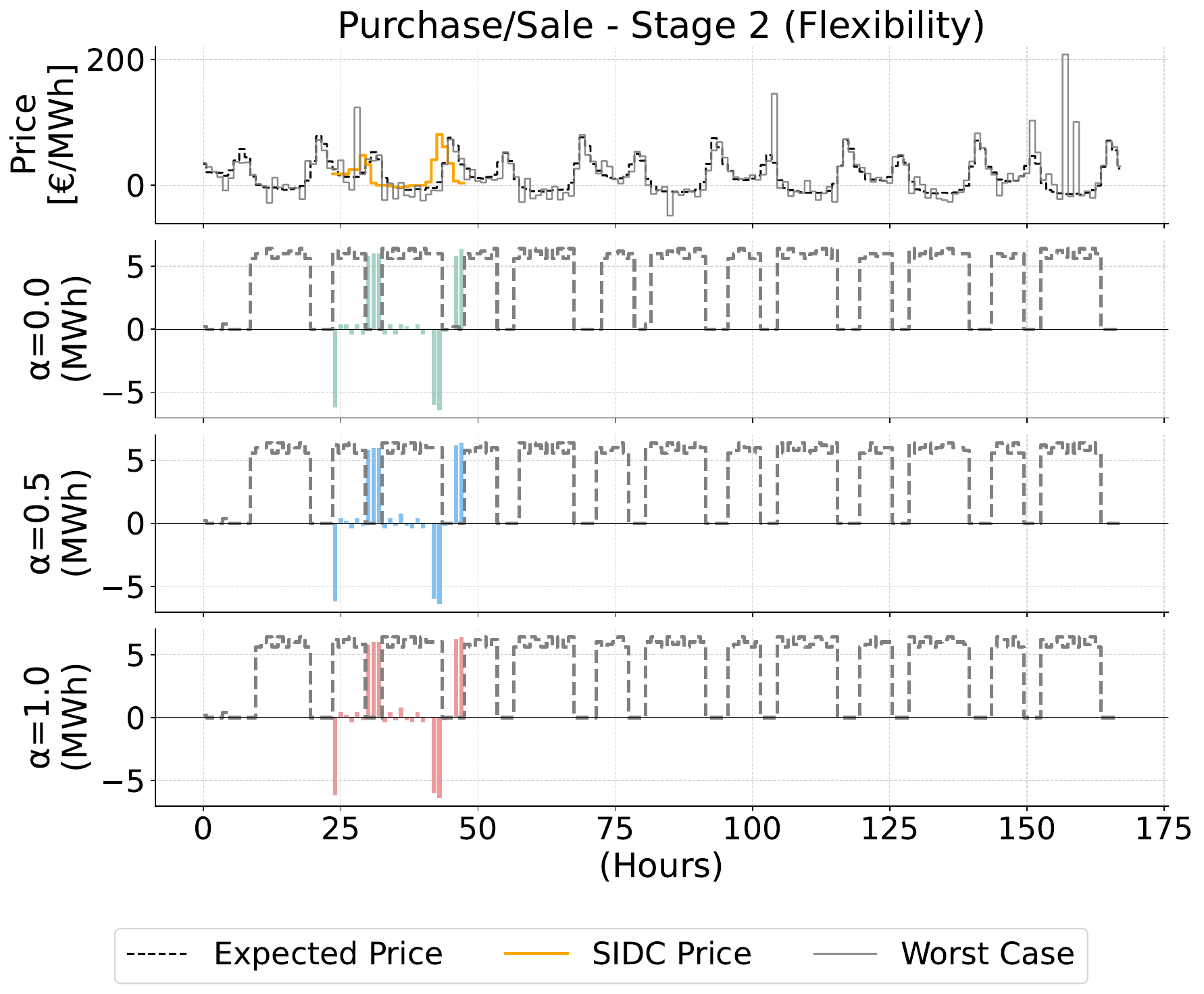}
\includegraphics[width=1\linewidth]{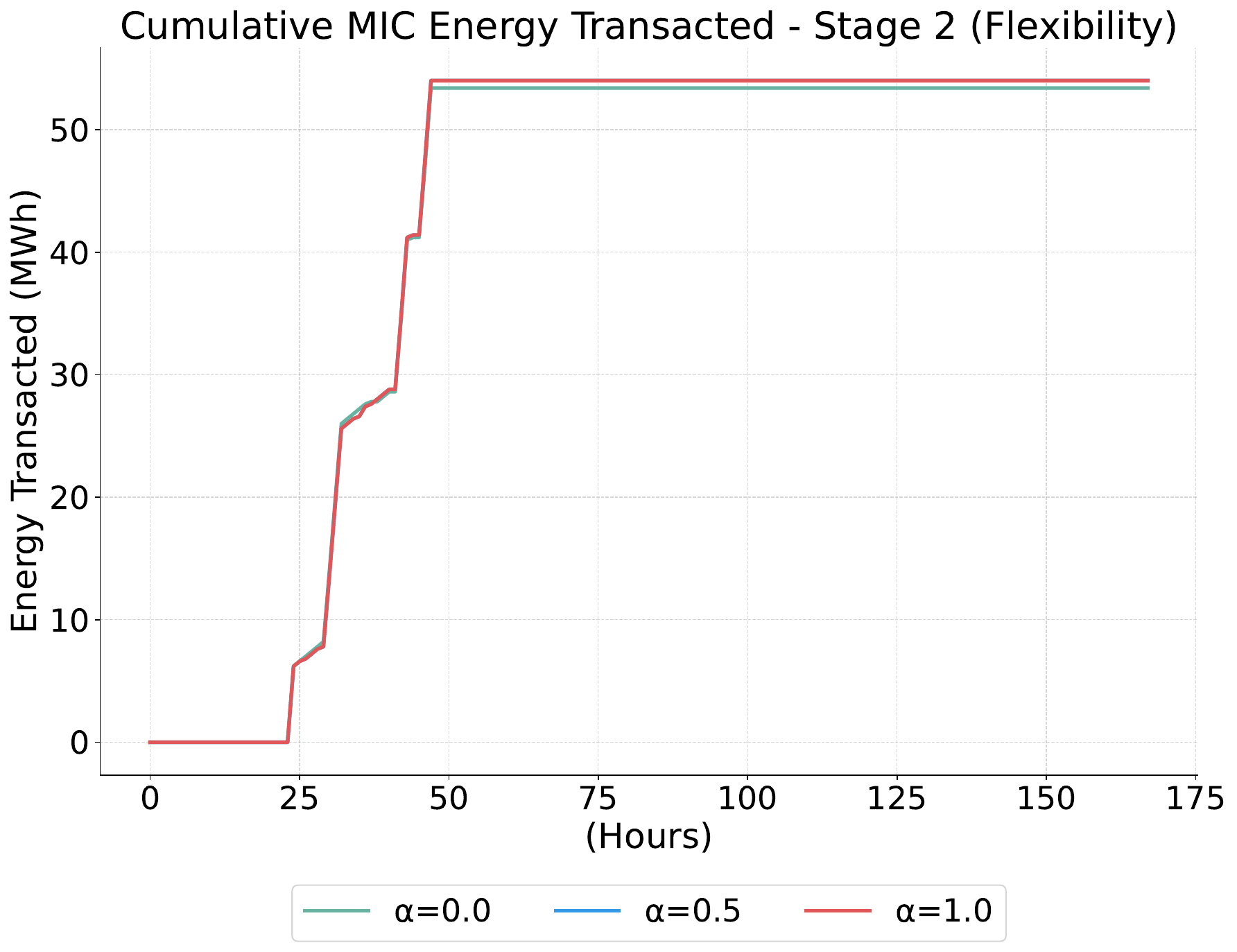}
\caption{Top: Hourly SIDC energy transacted compared to baseline day-ahead grid purchases across $\alpha$ levels. Bottom: Cumulative SIDC energy traded over time.}
\label{fig:sidc_trading_profiles}
\end{figure}

Figure~\ref{fig:sidc_trading_profiles} illustrates how the system engages with the SIDC market across different levels of risk aversion $\alpha$, focusing on the authorized trading window between hours 24 and 48. The top panel shows the hourly SIDC energy transactions under each $\alpha$ configuration, overlaid with the corresponding baseline grid purchases from Stage~1. The bottom panel plots the cumulative volume of SIDC energy traded over time (selling and buying), offering insight into the temporal dynamics of market participation.

A notable finding is the near-constancy of total traded volume: in all cases, the optimizer schedules approximately \SI{55}{MWh} of SIDC transactions. This consistency highlights the strategic value of SIDC engagement as a reliable flexibility mechanism, largely invariant to the chosen robustness setting. However, the temporal structure of these trades slightly varies with $\alpha$:

\begin{itemize}[left=0pt]
    \item \textbf{Risk-neutral strategies} ($\alpha = 1.0$) initiate trades earlier and with more concentrated bursts, exploiting arbitrage opportunities aggressively based on expected price differences.

    \item \textbf{Risk-averse strategies} ($\alpha = 0.0$) delay engagement or distribute transactions more evenly across the trading horizon, reflecting a cautious posture that avoids exposure to potential misalignment with baseline commitments.
\end{itemize}

These behaviors suggest that while the overall quantity of intraday market participation is robust to changes in risk preference, the optimizer’s temporal strategy—that is, when and how energy is traded—is sensitive to perceived uncertainty. This sensitivity stems from the indirect influence of the first-stage baseline schedule, which constrains feasible adjustments in the SIDC window.

It is also important to note that SIDC prices are treated as deterministic in the model’s second stage. Therefore, the optimizer’s decisions are driven by known price spreads between day-ahead and intraday markets rather than by SIDC price uncertainty. As a result, robustness influences SIDC participation indirectly—by shaping the flexibility of the baseline schedule and modulating the system’s willingness to deviate from it.

Finally, SIDC transactions operate in close coordination with BESS behavior. This co-activation strategy enables the system to buffer variability, optimize economic outcomes, and respond adaptively to evolving market conditions—reinforcing the value of integrated flexibility in risk-aware scheduling.

\subsubsection{Raw Mill Scheduling Dynamics}

\begin{figure}[ht]
\centering
\includegraphics[width=1\linewidth]{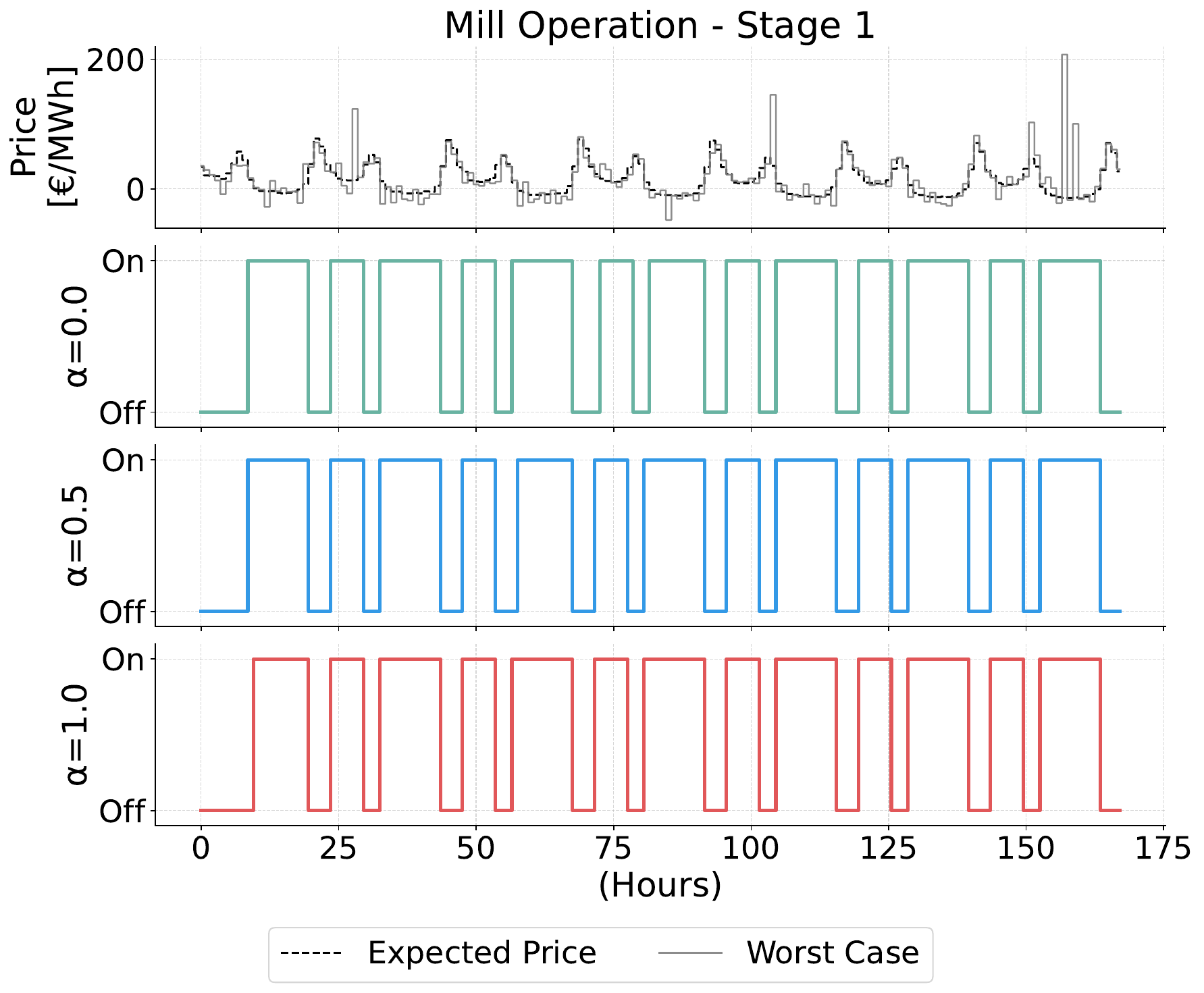}
\includegraphics[width=1\linewidth]{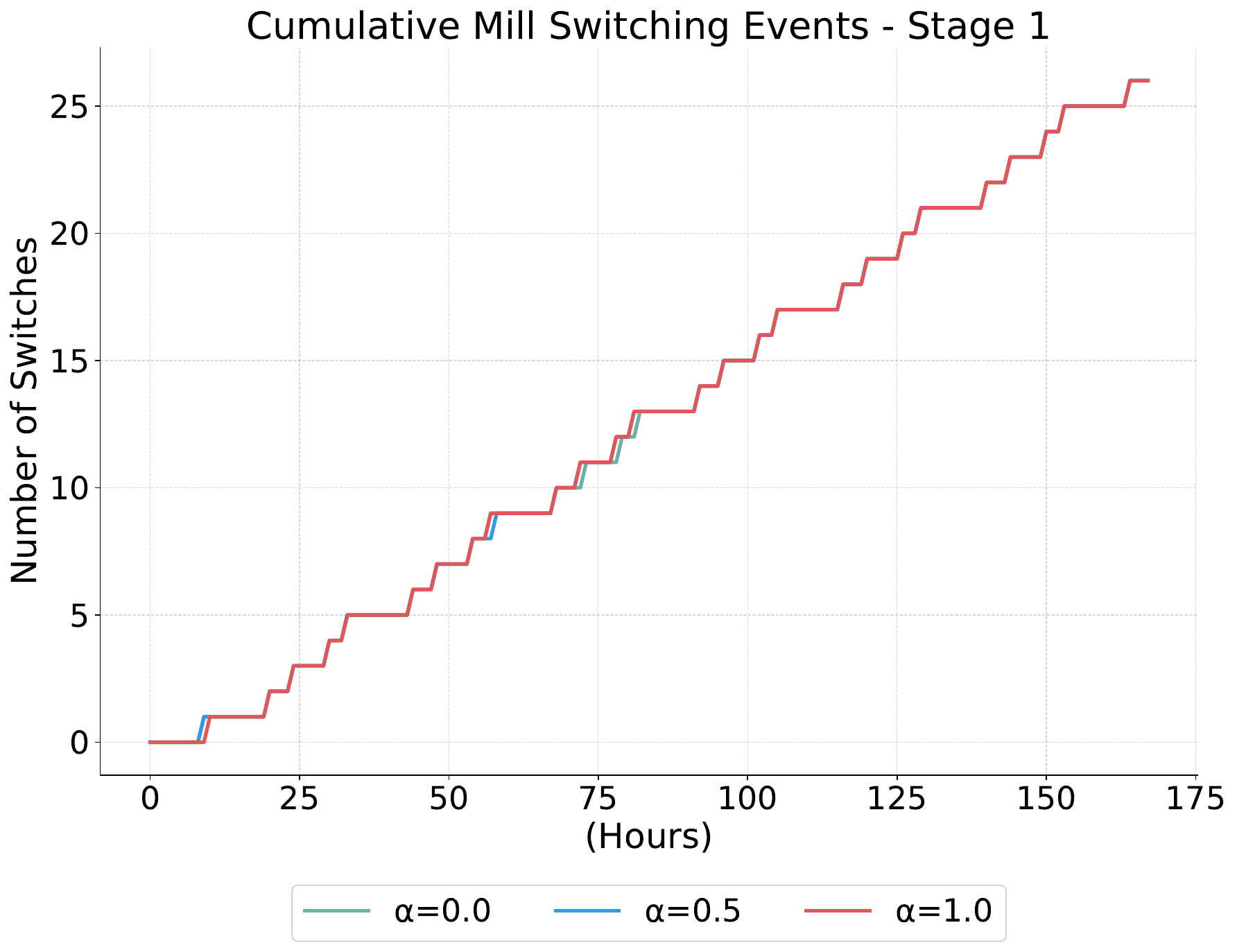}
\caption{Raw mill operational status (Top) and cumulative switching events (Bottom) in Stage~1 across $\alpha$ values.}
\label{fig:mill_operation_stage1}
\end{figure}

\begin{figure}[ht]
\centering
\includegraphics[width=1\linewidth]{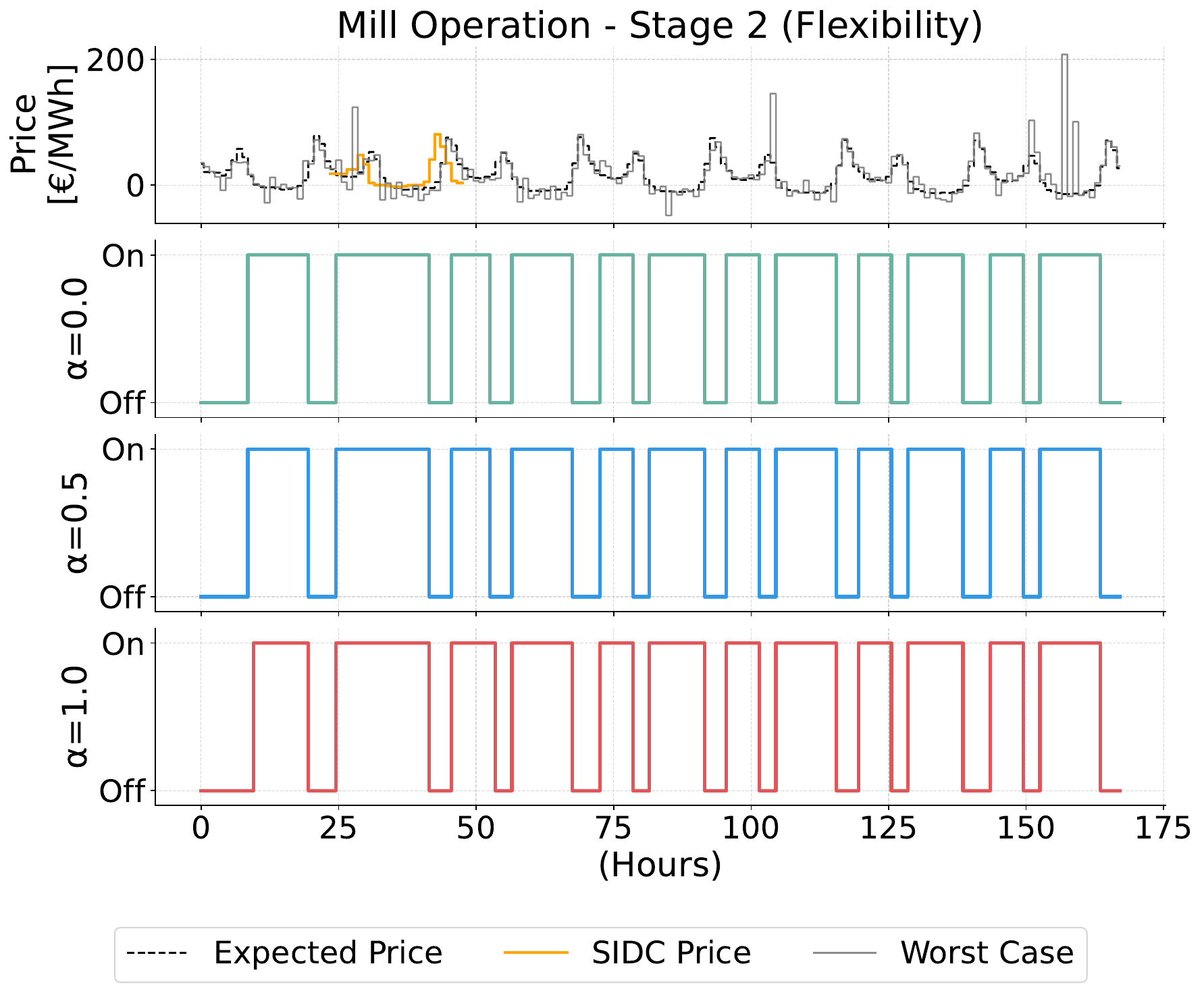}
\includegraphics[width=1\linewidth]{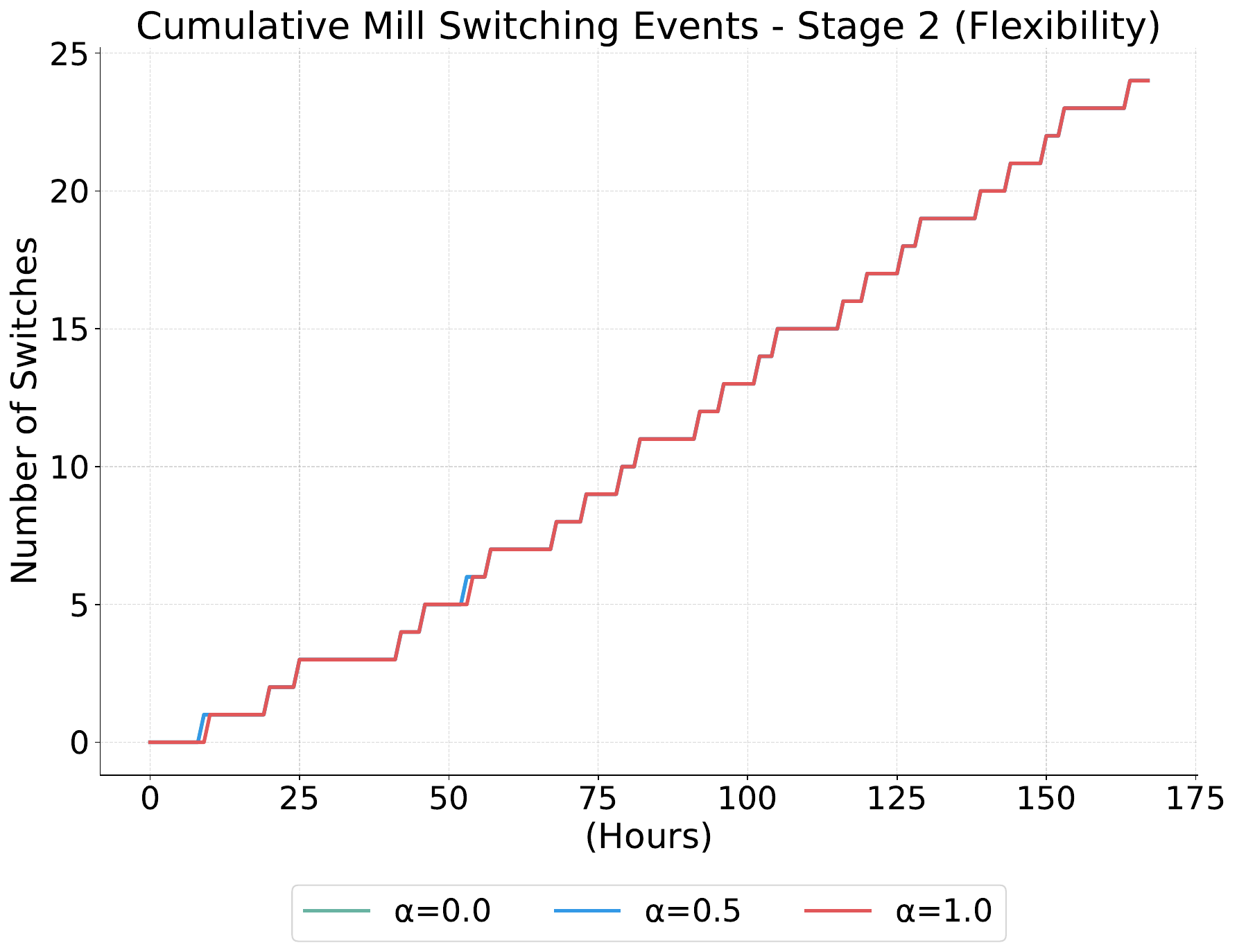}
\caption{Raw mill operational status (Top) and cumulative switching events (Bottom) in Stage~2 across $\alpha$ values. Slight smoothing is observed.}
\label{fig:mill_operation_stage2}
\end{figure}

Figures~\ref{fig:mill_operation_stage1} and~\ref{fig:mill_operation_stage2} illustrate the operation of the raw mill during Stage~1 (baseline scheduling) and Stage~2 (flexibility activation), respectively. As detailed in Section~\ref{sec:Methodology}, the binary scheduling decisions $Y_{k,t}$ governing the raw mill are non-adaptive variables—i.e., they are made prior to scenario realization and are held fixed across all uncertainty outcomes for a given $\alpha$ value. This structure mirrors the rigidity of industrial production commitments, which must often be determined in advance due to operational and safety constraints.

The top panels in each figure show the mill’s on/off status over time, superimposed on representative day-ahead price trajectories. Although the underlying schedule is scenario-invariant, different risk preferences ($\alpha$ values) lead to different pre-optimized patterns. Specifically, more conservative configurations (e.g., $\alpha=0.0$) tend to avoid operation during periods of high volatility or cost exposure, whereas risk-neutral settings (e.g., $\alpha=1.0$) are more opportunistic, scheduling production in line with expected arbitrage opportunities. Nonetheless, these differences are muted due to structural constraints—such as minimum uptime/downtime requirements and continuous kiln demand—that restrict the optimizer’s flexibility.

To quantify these variations, the bottom panels display the \textbf{cumulative number of switching events}, defined as the total count of on/off transitions over the planning horizon. This metric serves as a proxy for the schedule’s aggressiveness and mechanical stress on production assets.

In Stage~1 (Figure~\ref{fig:mill_operation_stage1}), the switching patterns are almost indistinguishable across different $\alpha$ values, reflecting the raw mill’s inherent scheduling rigidity. However, in Stage~2 (Figure~\ref{fig:mill_operation_stage2}), the availability of recourse via SIDC trading and battery flexibility slightly relaxes this rigidity. These auxiliary mechanisms absorb part of the system’s temporal imbalance, enabling the optimizer to meet energy needs with reduced reliance on mill switching. As a result, marginal reductions in switching frequency are observed for moderately conservative and risk-neutral configurations.

While the smoothing effect is modest, it underscores a key design insight: flexibility-enhancing assets can reduce operational wear and enable less aggressive production strategies. However, such benefits remain bounded by the physical and process limitations of industrial systems—particularly when core scheduling decisions must be scenario-agnostic.

This reinforces the necessity of distinguishing between \textbf{non-adaptive} and \textbf{adaptive} decision layers in robust industrial scheduling. While assets like BESS and SIDC trading offer dynamic response capabilities, large-scale production units such as the raw mill require conservative planning grounded in structural feasibility.

\subsubsection{Integrated Insights}

The operational analysis reveals a clear stratification in the behavior of flexible assets under uncertainty, shaped by the hierarchical structure of the robust optimization framework. Each asset exhibits a distinct role and degree of responsiveness depending on its place in the decision hierarchy and the system’s risk preferences:

\begin{itemize}[left=0pt]
    \item \textbf{Battery Energy Storage System}: Battery dispatch intensifies with decreasing risk aversion. At higher $\alpha$ values, the optimizer leverages expected price differentials more aggressively, leading to increased charge/discharge frequency and higher cumulative throughput. This confirms the BESS’s role as a versatile recourse mechanism, dynamically adapting to anticipated price dynamics to reduce procurement costs.

    \item \textbf{SIDC Market Participation}: SIDC trading emerges as a consistently activated strategy across all robustness levels. Although the total traded volume remains stable, the timing and temporal concentration of transactions vary with $\alpha$. Risk-neutral configurations engage earlier and more opportunistically, while conservative strategies adopt more cautious, distributed trading profiles. This demonstrates that SIDC participation is a low-regret flexibility lever whose value is modulated, rather than dictated, by risk tolerance.

    \item \textbf{Raw Mill Scheduling}: The raw mill’s operation remains invariant across all scenarios and $\alpha$ settings, reflecting its nature as a non-adaptive decision. Additionally, its scheduling is constrained by minimum uptime/downtime requirements and inventory dynamics. This rigid behavior highlights the necessity of pre-committing core production plans that are robust across potential future conditions.
\end{itemize}

Collectively, these observations underscore the importance of a modular and layered control architecture. The model distinguishes between:

\begin{itemize}[left=0pt]
    \item \textbf{Committed decisions} (e.g., production schedules), which are scenario-agnostic and must ensure feasibility across all uncertainty realizations;
    \item \textbf{Flexible decisions} (e.g., BESS dispatch and SIDC trades), which are scenario-responsive and optimized to exploit market variability and operational slack.
\end{itemize}

This design philosophy ensures robustness without sacrificing economic performance. It enables reliable non-adaptive planning while embedding adaptive layers that react to real-time conditions—a critical capability in volatile electricity markets. The result is a scheduling strategy that is not only resilient to forecast deviations but also capable of capturing value from emerging arbitrage opportunities.

Such hybrid decision frameworks are particularly well-suited to industrial settings, where operational inflexibility coexists with strategic flexibility assets. By aligning asset characteristics with decision levels, the proposed model provides a coherent and effective approach to short-term energy scheduling under uncertainty.

\section{Conclusions and Future Work}
\label{sec:conclusions}

This study has introduced a scenario-based robust optimization framework for short-term energy scheduling in electricity-intensive industrial plants, with a specific focus on managing uncertainty. The proposed methodology integrates hybrid scenario generation, two-stage MILP modeling, and a tunable risk-averse objective via the parameter $\alpha$, allowing planners to balance cost efficiency with operational robustness.

Applied to a real-world cement manufacturing facility, the framework demonstrates how coordinated use of industrial flexibility—through Battery Energy Storage System dispatch and intraday continuous market participation—can substantially reduce electricity costs while hedging against price volatility.

\subsection*{Key Contributions and Findings}

\begin{itemize}[left=0pt]
  \item \textbf{Hybrid scenario generation with preserved diversity:}  
  A three-step method combining ARIMA-based forecasting, stochastic residual sampling, and K-Means clustering produced a compact yet expressive scenario set. This enabled tractable robust optimization while preserving the statistical features of price uncertainty.

  \item \textbf{Quantification of the risk–cost trade-off:}  
  A sweep over the risk aversion parameter $\alpha$ highlighted the trade-off between worst-case protection and expected cost. Intermediate values (e.g., $\alpha = 0.5$–$0.8$) emerged as practical compromises, delivering robust yet cost-effective schedules.

  \item \textbf{Economic value of multi-asset flexibility:}  
  The integration of SIDC trading and BESS operation in Stage~2 consistently improved performance across all $\alpha$ levels. Flexibility reduced worst-case costs by up to 77\%, confirming its critical role in buffering against volatility.

  \item \textbf{Asset-specific behavioral insights:}  
  The framework uncovered differentiated asset behaviors under risk: BESS operation scaled with $\alpha$; SIDC trading remained uniformly active; and raw mill scheduling was invariant due to model design reflecting commitment schedule.

  \item \textbf{Superiority over deterministic scheduling:}  
  Compared to a deterministic baseline, robust strategies consistently achieved lower worst-case costs, with only modest increases in expected cost. This underscores the practical benefit of robustness in volatile market environments.

  \item \textbf{Validated feasibility under the horizon operation:}  
  All optimization instances were executed within a realistic horizon context using empirical data. The model satisfied all operational constraints, supporting its deployment for short-term industrial energy planning.

  \item \textbf{Scalability and tractability of the MILP formulation:}  
  Despite its multi-scenario nature, the model remained computationally manageable across tested configurations. No critical bottlenecks were encountered, and solution times scaled reasonably with the number of scenarios.
\end{itemize}

While effective, the current implementation relies on several simplifying assumptions that limit its generalizability. Only day-ahead electricity prices were treated as uncertain; other sources of variability—such as internal demand or PV generation—were held fixed or excluded from the simulations. The model also omits degradation effects, both in terms of BESS cycling and process wear due to frequent switching of production equipment. Furthermore, the plant is modeled as a price-taker, which may not hold in illiquid markets or under strategic bidding behavior. Finally, the case study is limited to a single industrial site, without accounting for network-level interactions or multi-site coordination. These assumptions were adopted to ensure tractability and clarity in the initial implementation, but future work should aim to relax them in order to improve realism and broaden applicability.

\subsection*{Future Research Directions}

The proposed framework opens multiple avenues for future extension:

\begin{itemize}[left=0pt]
  \item \textbf{Multi-source uncertainty:}  
  Incorporate variability in internal demand, PV generation, and other relevant market signals to enable a more comprehensive treatment of risk.

  \item \textbf{Multi-market optimization:}  
  Expand the scope to include coordinated participation in multiple electricity markets (e.g., day-ahead, SIDC, balancing reserves), capturing synergies and layered flexibility value.

  \item \textbf{Degradation-aware dispatch:}  
  Introduce lifetime models for storage and production equipment to reflect the long-term cost of frequent flexibility activation.

  \item \textbf{Integration with real-time control:}  
  Embed the robust MILP into rolling-horizon or Model Predictive Control architectures~\cite{parisio2016stochastic, oldewurtel2010use} to enable continual reoptimization with updated forecasts and prices.

  \item \textbf{Multi-objective formulations:}  
  Incorporate additional performance metrics—such as CO\textsubscript{2} emissions, renewable energy penetration, or resilience indices—to support sustainability-aligned decision-making.

  \item \textbf{Cross-site coordination:}  
  Extend the model to support scheduling across multiple industrial sites or aggregators, enabling collaborative flexibility provision and virtual power plant architectures.
\end{itemize}

\noindent
In conclusion, this work provides a practical and theoretically grounded approach for industrial energy consumers to manage uncertainty and leverage flexibility. The scenario-based robust optimization framework delivers resilient, economically efficient schedules while laying the groundwork for broader extensions in scale, complexity, and environmental scope.


\nomenclature[V]{$P_{b,t}$}{Power purchased from the grid at time $t$ [\unit{\mega\watt}].}
\nomenclature[V]{$P_{s,t}$}{Power exported to the grid at time $t$ [\unit{\mega\watt}].}
\nomenclature[V]{$P_{\mathrm{PV},t}$}{Power generated by the PV system at time $t$ [\unit{\mega\watt}].}
\nomenclature[V]{$P_{C,t}$, $P_{D,t}$}{Battery charging/discharging power at time $t$ [\unit{\mega\watt}].}
\nomenclature[V]{$P_{m,t}$}{Power traded in the SIDC market at time $t$ [\unit{\mega\watt}].}
\nomenclature[V]{$Y_{k,t}$}{Binary on/off status of production unit $k$ at time $t$.}
\nomenclature[V]{$I_{i,t}$}{Inventory level in silo $i$ at time $t$ [\unit{\tonne}].}
\nomenclature[V]{$\mathrm{SOC}_t$}{Battery state of charge at time $t$ [\unit{\mega\watt\hour}].}
\nomenclature[V]{$\Phi$}{Total electricity procurement cost [\unit{\euro}].}
\nomenclature[V]{$\Phi^*$}{Electricity cost of the baseline schedule [\unit{\euro}].}
\nomenclature[V]{$\Phi^\dagger$}{Electricity cost after flexibility activation [\unit{\euro}].}
\nomenclature[V]{$\Delta \Phi$}{Flexibility profit: cost reduction due to SIDC participation [\unit{\euro}].}
\nomenclature[V]{$\Phi^{\mathrm{BL},s}$}{Baseline cost under scenario $s$ [\unit{\euro}].}
\nomenclature[V]{$\Phi^{\mathrm{FLX},s}$}{Flexible cost under scenario $s$ [\unit{\euro}].}

\nomenclature[C]{$\pi_{b,t}$}{Day-ahead market price at time $t$ [\unit{\euro\per\mega\watt\hour}].}
\nomenclature[C]{$\pi_{m,t}$}{SIDC market price at time $t$ [\unit{\euro\per\mega\watt\hour}].}
\nomenclature[C]{$\pi_{S,i,t}$}{Inventory holding cost coefficient for silo $i$ at time $t$ [\unit{\euro\per\tonne}].}
\nomenclature[C]{$\pi_U$}{Battery usage penalty coefficient [\unit{\euro\per\mega\watt\hour}].}
\nomenclature[C]{$P_k$}{Rated power of production unit $k$ [\unit{\mega\watt}].}
\nomenclature[C]{$\Pi_k$}{Production throughput of unit $k$ [\unit{\tonne\per\hour}].}
\nomenclature[C]{$I_{i,\min}$, $I_{i,\max}$}{Minimum and maximum inventory in silo $i$ [\unit{\tonne}].}
\nomenclature[C]{$C_{\max}$}{Maximum battery capacity [\unit{\mega\watt\hour}].}
\nomenclature[C]{$\mathrm{DoD}$}{Depth of discharge of battery [dimensionless].}
\nomenclature[C]{$P_C^{\max}$, $P_D^{\max}$}{Maximum battery charge/discharge rate [\unit{\mega\watt}].}
\nomenclature[C]{$P_b^{\max}$}{Maximum power purchased from the grid [\unit{\mega\watt}].}
\nomenclature[C]{$M_k^{\mathrm{ON}}$, $M_k^{\mathrm{OFF}}$}{Minimum on/off time for unit $k$ [\unit{\hour}].}
\nomenclature[C]{$\alpha$}{Risk aversion parameter in the robust objective [dimensionless].}
\nomenclature[C]{$\rho_s$}{Probability of scenario $s$ [dimensionless].}
\nomenclature[C]{$\Delta t$}{Time step duration [\unit{\hour}].}
\nomenclature[C]{$\tau_1$, $\tau_2$}{SIDC trading window limits [\unit{\hour}].}
\nomenclature[C]{$LC1$}{Maximum SIDC trading capacity [\unit{\mega\watt}].}

\nomenclature[P]{$t$}{Index for time periods.}
\nomenclature[P]{$s$}{Index for uncertainty scenarios.}
\nomenclature[P]{$k$}{Index for production units.}
\nomenclature[P]{$i$}{Index for silos.}
\nomenclature[P]{$\mathcal{T}$}{Set of time periods.}
\nomenclature[P]{$\mathcal{K}$}{Set of production units.}
\nomenclature[P]{$\mathcal{S}$}{Set of silos.}
\nomenclature[P]{$S$}{Set of uncertainty scenarios.}

\printnomenclature

\printcredits

\section*{Acknowledgments}
We appreciate the assistance of Fortia Energía for providing the related information on the Industrial Case Study.

\section*{Financial disclosure}
This research has been supported by funding from the European Commission under a project related to industrial flexibility and energy transition.

\section*{Conflict of interest}
The authors have no relevant financial or non-financial interests to disclose.

\section*{Data Availability}
All electricity market data used in this study are publicly available from the Spanish System Operator Information System (e·sios) platform, operated by Red Eléctrica de España (REE). The data—including day-ahead and intraday electricity prices—can be accessed at: \url{https://www.esios.ree.es/en}.

\section*{Declaration of generative AI and AI-assisted technologies in the writing process}
During the preparation of this work the authors used \textit{ChatGPT} (developed by OpenAI) to enhance readability and improve language clarity. After using this tool, the authors reviewed and edited the content as needed and take full responsibility for the content of the published article.

\bibliographystyle{cas-model2-names}

\bibliography{cas-refs}

\end{document}